\documentstyle[nato,epsf]{crckapb} 

\newbox\grsign \setbox\grsign=\hbox{$>$} \newdimen\grdimen \grdimen=\ht\grsign
\newbox\simlessbox \newbox\simgreatbox
\setbox\simgreatbox=\hbox{\raise.5ex\hbox{$>$}\llap
     {\lower.5ex\hbox{$\sim$}}}\ht1=\grdimen\dp1=0pt
\setbox\simlessbox=\hbox{\raise.5ex\hbox{$<$}\llap
     {\lower.5ex\hbox{$\sim$}}}\ht2=\grdimen\dp2=0pt

\def\simless{\mathrel{\copy\simlessbox}}
\def\etal{{\it et al.~}}
\def\kms{\ifmmode {\rm km~s^{-1}}\else ${\rm km~s^{-1}}$\fi}
\def\degree{\ifmmode {^\circ}\else {$^\circ$}\fi}

\def\rstar{\ifmmode {R_{\star}}\else $R_{\star}$\fi}
\def\rsun{\ifmmode {\rm R_{\odot}}\else $\rm R_{\odot}$\fi}
\def\rsunsq{\ifmmode {\rm R_{\odot}^2}\else $\rm R_{\odot}^2$\fi}
\def\mstar{\ifmmode {M_{\star}}\else $M_{\star}$\fi}
\def\lstar{\ifmmode {L_{\star}}\else $L_{\star}$\fi}
\def\tstar{\ifmmode {T_{\star}}\else $T_{\star}$\fi}
\def\msun{\ifmmode {\rm M_{\odot}}\else $\rm M_{\odot}$\fi}
\def\msunyr{\ifmmode {\rm M_{\odot}\,yr^{-1}}\else $\rm M_{\odot}\,yr^{-1}$\fi}\def\mdot{\ifmmode {\dot{M}}\else $\dot{M}$\fi}
\def\mssunyr{\ifmmode {\rm M_{\odot}^2\,yr^{-1}}\else $\rm M_{\odot}^2\,yr^{-1}$\fi}\def\mdot{\ifmmode {\dot{M}}\else $\dot{M}$\fi}
\def\lsun{\ifmmode {\rm L_{\odot}}\else $\rm L_{\odot}$\fi}
\def\lbol{\ifmmode {L_{bol}}\else $L_{bol}$\fi}
\def\teff{\ifmmode {T_{eff}}\else $T_{eff}$\fi}
\def\ne{\ifmmode {n_{e}}\else $n_{e}$\fi}
\def\te{\ifmmode {T_{e}}\else $T_{e}$\fi}
\def\cm3{\ifmmode {\rm cm^{-3}}\else $\rm cm^{-3}$\fi}
\def\emm{\ifmmode {n_e^2 V}\else $n_e^2 V$\fi}
\def\gcm3{\ifmmode {\rm g~cm^{-3}}\else $\rm g~cm^{-3}$\fi}
\def\ergg{\ifmmode {\rm erg~g^{-1}}\else $\rm erg~g^{-1}$\fi}
\def\ergs{\ifmmode {\rm erg~s^{-1}}\else $\rm erg~s^{-1}$\fi}
\def\ecs{\ifmmode {\rm erg~cm^{-2}~s^{-1}}\else $\rm erg~cm^{-2}~s^{-1}$\fi}
\def\mum{\ifmmode {\rm \mu {\rm m}}\else $\rm \mu {\rm m}$\fi}
\def\nh3{\ifmmode {\rm NH_3}\else $\rm NH_3$\fi}
\def\arcsec{\ifmmode ^{\prime \prime}\else $^{\prime \prime}$\fi}

\def\inch{\ifmmode ^{\prime \prime}\else $^{\prime \prime}$\fi}
\def\arcmin{\ifmmode ^{\prime}\else $^{\prime}$\fi}

\def\lfl{\ifmmode {\lambda F_{\lambda}}\else $\lambda F_{\lambda}$\fi}
\def\lFl{\ifmmode {\lambda F_{\lambda}}\else $\lambda F_{\lambda}$\fi}
\def\footspace{\baselineskip=10pt}
\def\singlespace{\baselineskip=12pt}

\begin{opening}
\title{ACCRETION DISKS AND ERUPTIVE PHENOMENA}

\author{Scott J. Kenyon}
\institute{Smithsonian Astrophysical Observatory\\
           60 Garden Street, Cambridge, MA 02138 USA}

\end{opening}

\runningtitle{FILE}

\begin{document}


\section{Introduction}

In the 1700's, Immanuel Kant and the Marquis de Laplace proposed 
that the solar system collapsed from a gaseous medium of roughly 
uniform density (\cite{kan55}, \cite{lap96}).  A flattened gaseous 
disk -- the protosolar nebula -- formed out of this cloud.  The Sun 
contracted out of material at the center of the disk, while 
the planets condensed in the outer portions.  Despite its simplicity, 
this model suffered from the {\it angular momentum problem} inherent 
to star formation: a cloud of gas and dust with the diameter of 
the solar system and the mass of the Sun has too much angular 
momentum to collapse to the Sun's present size.  It took 150 years 
to solve this problem. C. von Weisz\"acker worked out the basic 
physics of a viscous accretion disk, building on previous realizations
that a turbulent viscosity could move material inwards and angular 
momentum outwards through the protosolar nebula (\cite{von43}, 
\cite{von48}).  With L\"ust's steady-state solution (\cite{lus52}),
Kant's nebular
hypothesis no longer suffered from the angular momentum problem
and became the leading model for solar system formation.

Disk accretion is now known to power a wide range of astronomical 
sources.  Starting with Kuiper's landmark study of $\beta$ Lyr, 
viscous accretion disks have become central to our understanding 
of interacting binaries, where one star fills its inner Lagrangian
surface and transfers matter into a disk surrounding a smaller 
companion (\cite{kui41}).  A disk surrounding a supermassive 
black hole is widely believed to drive the phenomena observed
in active galactic nuclei (e.g., \cite{ulr97}).  Accretion disks
remain an important issue in star and planet formation, as 
described throughout this volume.

One striking feature of recent observations is the common phenomena 
observed in different accreting systems.  All disks vary in brightness.  
These fluctuations range from rapid flickering on time scales 
as short as a few milliseconds up to long-lived eruptions 
with durations of decades or centuries.
Most accreting systems lose mass; practically all disks that lose
mass eject material in a well-collimated jet (\cite{liv97}).  
These common phenomena occur in systems with luminosities
ranging from much less than 1 \lsun~in some interacting binaries 
up to $10^{11}~\lsun$ in active galactic nuclei.

\vskip 2ex
\epsffile{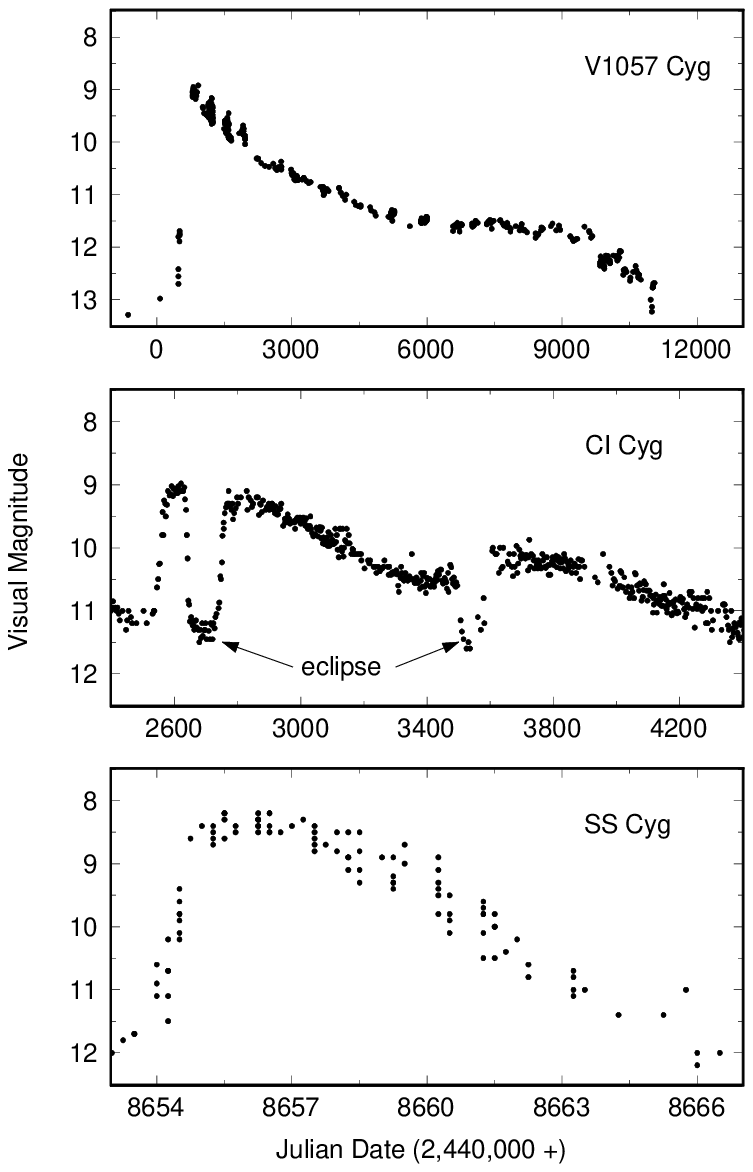}

\footnotesize
\footspace

\vskip -4ex
\noindent
Figure 1 - Light curves for SS Cyg, a cataclysmic binary;
CI Cyg, a symbiotic binary; and V1057 Cyg, an FU Ori variable star.
Despite a different size scale in each system, the light curves have 
common features that indicate a similarity in their underlying physics.

\vskip 4ex
\normalsize
\singlespace

As an example of these common phenomena, Figure 1 shows 
light curves for SS Cyg -- a cataclysmic binary with an orbital 
period of $P_{orb}$  = 6.6 hr (\cite{can98}); CI Cyg -- a symbiotic binary 
with $P_{orb}$ = 855 days (\cite{ken86}, \cite{bel92}); 
and V1057 Cyg -- an FU Ori variable that is apparently a single star
(\cite{ken91a}).  The 3--5 mag eruptions of these systems 
occur when the accretion rate through the disk increases. Aside from 
the deep eclipses in CI Cyg, the three light curves have common
features, with a rapid rise, a brief interval at maximum, and a 
long decay.  All three systems resemble A- or F-type stars at 
maximum light. This spectrum cools as the brightness fades.  
Substantial mass loss is associated with each system; the mass loss
probably increases with the rise in brightness and decreases as the
brightness declines. These similarities suggest that the same 
physical processes govern the evolution of disks ranging in size 
from 0.1 \rsun~to several tens of AU.

\vskip -2ex
\begin{table}[htb]
\begin{center}
\caption{Types of disk systems}
\begin{tabular}{lccc}
\hline 
System & Distance (pc) & Angular size (arcsec) & $N_{res}$ \\
\hline
BH + MS & 1000 & $\simless 10^{-5}$ & 100 \\
NS + MS & 1000 & $\simless 10^{-5}$ & 100 \\
WD + MS & 100~ & $\simless 10^{-3}$ & 100+ \\
MS + SG & 100~ & $\simless 10^{-3}$ & 10+ \\
MS + RG & 1000 & $\simless 10^{-3}$ & 10--100 \\
\\
$\beta$ Pic & 10+ & 1--10 & 10--100 \\
Young star  & 100+ & 1--2 & 10--20 \\
\\
AGN & $10^8$ & $\simless 10^{-3}$ & 1--10 \\
\hline
\end{tabular}
\end{center}
\end{table}

The common features of accreting systems provide good tools 
to test physical models of disks.  Table 1 summarizes the main 
types of disk-accreting systems known today and provides simple 
comparisons of their angular sizes and the number of
resolution elements across the face of the disk, $N_{res}$,
that can be derived from modern observations.  The first set 
of systems includes
(i) compact binaries containing a disk surrounding a black hole (BH), 
a neutron star (NS), or a white dwarf star (WD), and 
(ii) wider binaries with a disk surrounding a main sequence star.
A low mass main sequence star usually feeds the disk in compact binaries;
a more evolved subgiant or red giant star feeds the disk in the wider
binaries.  In both types of binaries, the angular size of the disk
is small compared to the resolution possible with current ground-based 
or space-based observatories.  However, eclipses of the disk by the 
secondary star can be used to divide the disk into many resolution 
elements and to map out the physical structure of the disk with a 
few simplifying assumptions (\cite{ho85a}, \cite{ho85b}).  
The dusty disks in the second group of systems are large enough to
image directly at optical and radio wavelengths, although few
systems have been mapped in much detail (\cite{smi84}, \cite{jay98}).  
The final group of active galactic nuclei have only recently been 
mapped with VLBI techniques; $N_{res}$ will probably increase as
these techniques become more sophisticated (\cite{her98}).

\section{Steady Disks}

Accretion disks work to transport mass radially inwards and 
angular momentum radially outwards.  To understand how a disk
manages this feat, consider a thin ring with two adjacent annuli 
at distances $r_1$ and $r_2$ from a central star (Figure 2). 
Material in these annuli orbits the central star with velocities, 
$v_1$ and $v_2$. The velocity difference between the two annuli,
$v_1 - v_2 > 0$, produces a frictional force that attempts to equalize 
the two orbital velocities.  The energy lost to friction heats 
the annuli; some disk material then moves inwards to conserve
total energy.  This inward mass motion increases the angular momentum
of the ring; some disk material moves outwards to conserve
angular momentum.  Energy and angular momentum conservation thus 
lead to an expansion of the ring in response to frictional heating.
The ring eventually expands into a disk, which generates heat 
(and radiation) at a level set by the accretion rate.

\vskip -1ex
\epsfxsize=4.0in
\epsffile{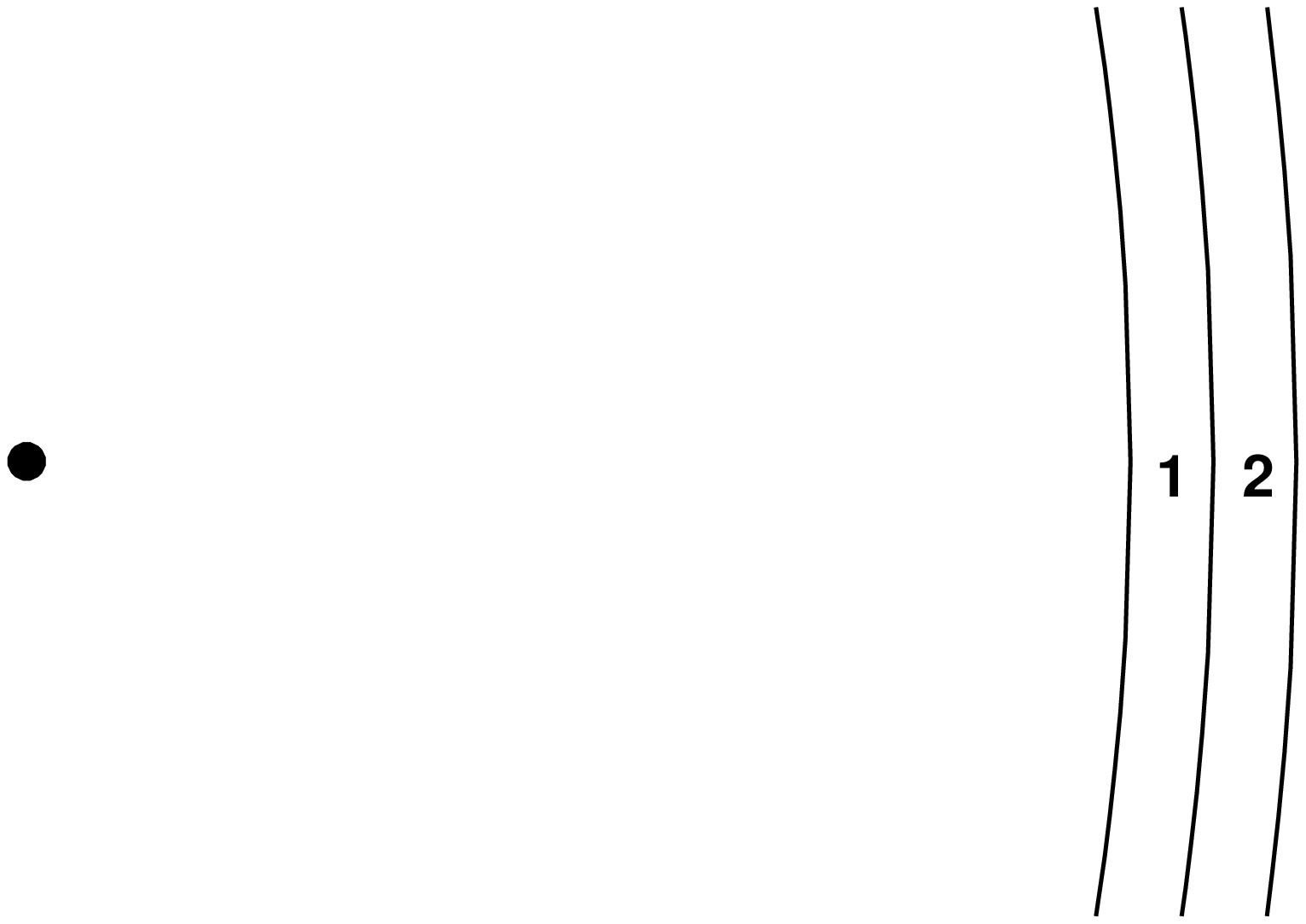}

\footnotesize
\footspace

\vskip -3ex
\noindent{Figure 2 - Schematic view of two adjacent annuli in a disk 
surrounding a compact star.  Annulus `1' lies inside annulus `2'
and orbits the star (filled circle) at a higher velocity,
$v_1 > v_2$.}

\singlespace
\normalsize

\subsection{Accretion Luminosities and Temperatures}

Figure 3 shows the standard disk geometry.  In a {\it steady} disk, 
material drifts radially inward at a constant rate, $\mdot$.  
For an infinite disk, the total luminosity generated by accretion
is $L_{acc}$ = $G \mstar \mdot/\rstar$, which is

\begin{equation}
L_{acc} = 314~\lsun \left ( \frac{\mstar}{1~\msun} \right )
\left ( \frac{\mdot}{10^{-5}~\msunyr} \right )
\left ( \frac{1~\rsun}{\rstar} \right ) 
\end{equation}

\epsfxsize=5.in
\epsffile{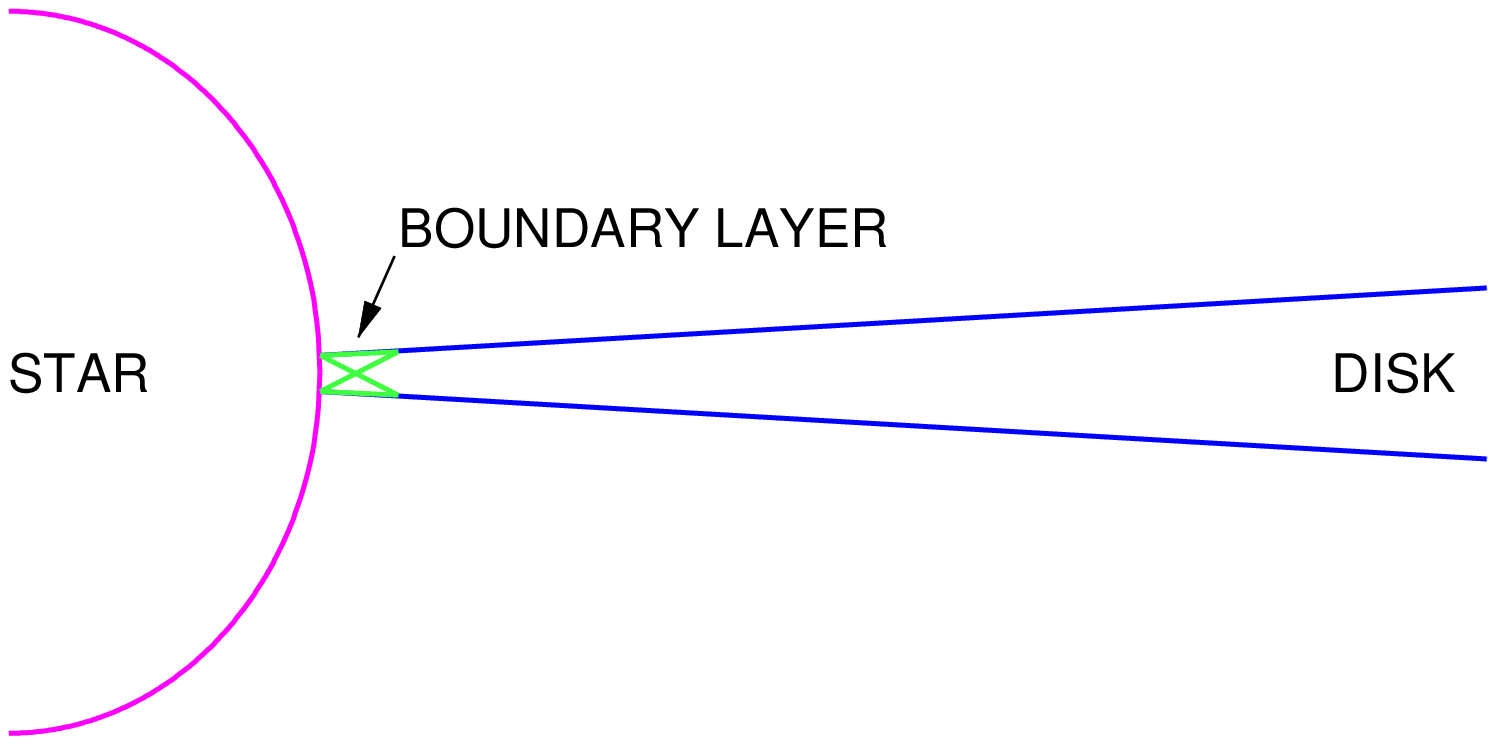}

\footnotesize
\footspace

\vskip -23ex
\noindent{Figure 3 - Schematic view of a disk that extends to the 
stellar photosphere.  Disk material drifts radially inward until it
reaches the boundary layer, where the rotational velocity slows to
match the stellar rotational velocity.}

\singlespace
\normalsize

\vskip 3ex
\noindent 
in familiar units.  As material drifts inwards, half of this accretion
energy is radiated by the disk, 

\begin{equation}
L_{disk} = \frac{1}{2} L_{acc} ~ .
\end{equation}

\noindent
The other half of the accretion energy becomes kinetic energy of orbital 
motion around the central star.  In most cases, disk material drifts 
inwards until it reaches the stellar photosphere (Figure 3).  
At this point, disk material with an angular velocity of 
$\Omega = (G \mstar/R_{\star}^3)^{1/2}$
must slow down to match the stellar angular velocity, $\Omega_{\star}$.  
This transition occurs in the ``boundary layer,'' a narrow ring with
a radial size,

\begin{equation}
R_{bl,d} \sim h_{bl}^2/\rstar \ll \rstar,
\end{equation}

\noindent
where $h_{bl}$ is the local scale height ($h_{bl}/\rstar \ll 1$; 
\cite{pri77}, \cite{pri79}, \cite{reg83}, \cite{pap86}, \cite{kle87}, 
\cite{reg88}, \cite{kle91}).
The energy lost in the boundary layer is the difference between
the rotational energy per unit mass in the disk and the rotational
energy per unit mass in the central star:

\begin{equation}
L_{BL} = \frac{1}{2} L_{acc} ~ \left 
( \frac{\omega_{disk}^2 - \omega_{\star}^2}{\omega_{disk}^2} \right ) ~ .
\end{equation}

In some accreting systems, the magnetic field of the central star
is strong enough to truncate the disk at a radius, $R_m > \rstar$.
A rough estimate of the truncation radius is the Alfv\'en radius,

\epsfxsize=5.in 
\epsffile{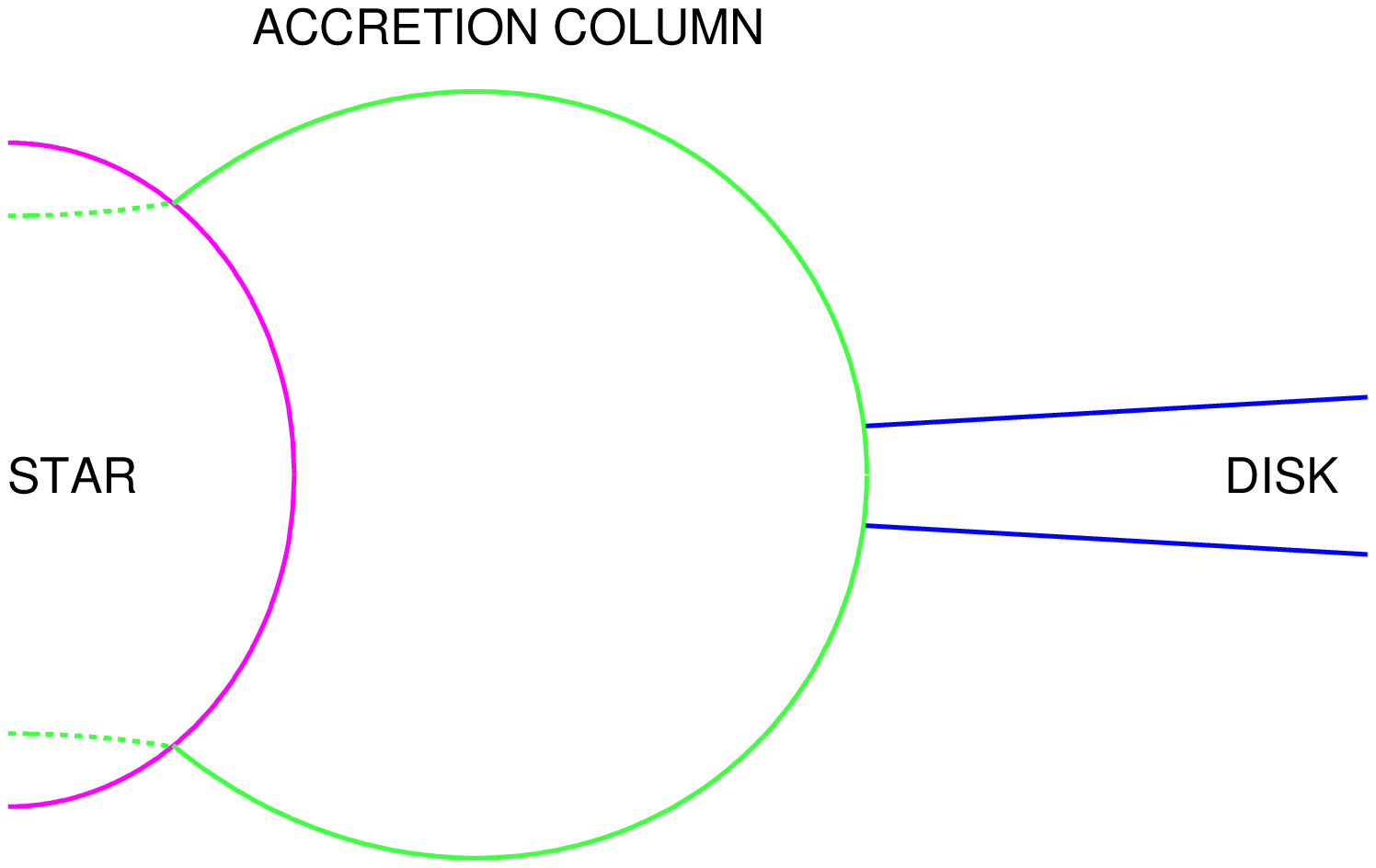}
 
\footnotesize
\footspace

\vskip -15ex
\noindent{Figure 4 - Schematic view of a disk truncated by a strong
dipolar magnetic field.  The inner radius of the disk lies close to 
the corotation radius, where the disk and the stellar photosphere have
the same angular velocity.  Material falls from the inner disk onto the 
star along the magnetic field lines and produces two circular rings surrounding 
the magnetic axis.}

\vskip 4ex
\normalsize
\singlespace

\begin{equation}
R_m \approx R_A \approx \left ( \frac{\mu_{\star}^4}{2 G \mstar \mdot^2} \right )^{1/7} 
\end{equation}

\noindent
where $\mu_{\star}$ is the magnetic dipole moment of the central star.
More rigorous estimates of $R_m$ depend on the magnetic field geometry
(\cite{lip78}, \cite{gho79}, \cite{lip80}, \cite{cam87}, \cite{kon91}, 
\cite{yi94}, \cite{arm95}, \cite{li96}, \cite{wan97}).
The disk then has a total luminosity

\begin{equation} 
L_{disk} = \frac{1}{2} L_{acc} ~ \left ( \frac{\rstar}{R_m} \right )
\end{equation}

\noindent
In this geometry, disk material falls onto the star along magnetic field 
lines (Figure 4).
The infalling gas shocks and produces two rings at stellar latitudes,
$\pm b$, if the magnetic and rotational axes are parallel (\cite{gho79},
\cite{kon91}). These rings are compressed into tilted ellipses if the 
magnetic axis is inclined with respect to the rotational axis, as
expected in most systems (\cite{ken94}, \cite{mah98}). In both cases, 
the luminosity of one ring is

\begin{equation}
L_{ring} = \frac{1}{2} L_{acc} ~ \left ( 1 - \frac{1}{2} \frac{\rstar}{R_{mag}} \right )
\end{equation}

To derive the temperature structure of the disk and the boundary layer
or magnetic accretion ring, we again use conservation of energy.  The
disk temperature assumes an approximate balance between the energy 
lost from blackbody radiation and the torque due to frictional heating 
from mass motion through a distance $\Delta R$ in a gravitational 
potential: 

\begin{equation}
G \mstar \mdot \left ( \frac{1}{R} - \frac{1}{R + \Delta R} \right ) \approx
2 \cdot 2 \pi R \Delta R ~ \sigma T_d^4 
\end{equation}

\noindent
For $\Delta R \ll R$, this expression is

\begin{equation} 
G \mstar \mdot \frac{\Delta R}{R^2} \approx 4 \pi R \Delta R \sigma T_d^4   
\end{equation} 

\noindent
which results in

\begin{equation}
T_d \approx \left ( \frac{G \mstar \mdot}{4 \pi \sigma R^3} \right )^{1/4} \propto R^{-3/4} 
\end{equation}

The exact temperature of a steady disk depends on how the torque transports
energy through the disk.  Standard models assume that frictional heating 
vanishes at the stellar photosphere, which leads to an extra term in 
the energy equation (8) above (\cite{lyn74}, \cite{sha73}).  The temperature 
is then (\cite{bat74}):

\begin{equation}
T_d(R) = T_{acc} \left ( \frac{R}{\rstar} \right )^{-3/4}
\left ( 1 - \sqrt{\rstar/r} \right )^{1/4}
\end{equation}

\noindent 
where

\begin{equation}
T_{acc} = 13000~{\rm K} \left ( \frac{\mstar \mdot}{10^{-5}~\mssunyr} \right )^{1/4}
\left ( \frac{1~\rsun}{\rstar} \right )^{3/4} .
\end{equation}

\noindent
This expression for the temperature vanishes at the stellar surface,
because the frictional heating vanishes.  The temperature reaches
a maximum at $R/\rstar$ = 49/36 (\cite{bat74}):

\begin{equation}
T_{max} = 6500~{\rm K} \left ( \frac{\mstar \mdot}{10^{-5}~\mssunyr} \right )^{1/4}
\left ( \frac{1~\rsun}{\rstar} \right )^{3/4}
\end{equation}

\noindent
Different boundary conditions at the stellar surface yield slightly 
different, non-analytic, temperature laws (\cite{pop91a}, \cite{pop91b}).
These 
results are usually close to the standard temperature law, equation (11).

The temperature of the boundary layer depends on the mass accretion
rate.  At high \mdot, the region is optically thick (\cite{lyn74},
\cite{tyl81}). Energy generated in the thin ``dynamical boundary layer'' 
(equation 3) should diffuse over a broader ring with a size 
comparable to the thermal scale height, $h_{\rm bl}$ (\cite{pap86}).
The scale height is smaller than the stellar radius in most cases.
This ``thermal boundary layer'' is then much larger than the dynamical
boundary layer and has a temperature 4--5 times larger than the
maximum disk temperature (\cite{ken84}).  For low $\mdot$, 
the boundary layer is optically thin and cannot radiate energy 
efficiently ({\cite{pri77}, \cite{pri79}).  The thermal scale 
height is then comparable to the stellar radius, and the 
boundary layer resembles a hot stellar corona (\cite{pri77}, 
\cite{pri79}).  The details of boundary layer structure are sensitive
to properties of the stellar photosphere and the viscosity mechanism,
but these general considerations hold for many physical settings
(\cite{ber92}, \cite{ber93}, \cite{pop93}, \cite{hu95a}, 
\cite{hu95b}, \cite{god95}, \cite{god96}, \cite{pop96}, 
\cite{ida96}, \cite{kle96}, \cite{kle97}).

The temperature of magnetic accretion rings also depends on
geometry.  If the magnetic axis is aligned with the rotational axis,
the rings have equal luminosities and temperatures.  The temperature 
is also constant with azimuth along each ring (\cite{mah98}).  The two 
rings have unequal luminosities when the magnetic axis is tilted with
respect to the rotational axis, as needed to produce the observed
light variations in magnetic cataclysmic variables and T Tauri stars 
(\cite{cro90}, \cite{ken94}).  The luminosity also varies with azimuth 
along each ring (\cite{mah98}).  
In the aligned case, the infalling matter shocks above the stellar 
surface and forms identical optically thick accretion columns that 
radiate in the Balmer and Paschen continua (\cite{lam98}, \cite{cal99}).
The surface area of this emission is comparable to the surface area
of the boundary layer, so the temperature is still 3--5 times larger
than the maximum disk temperature (\cite{cal99}).  The accretion
columns are probably quite different in non-aligned cases, but 
their structure has not been addressed.

\vskip -2ex
\begin{table}[htb]
\begin{center}
\caption{Accretion and nuclear energies}
\begin{tabular}{lcc}
\hline 
Object & $\epsilon_g$ (\ergg) & $\epsilon_{nuc}$ (\ergg) \\
\hline
T Tauri star       & $5 ~ \times ~ 10^{14}$ & $5 \times 10^{13}$ \\
main sequence star & $2 ~ \times ~ 10^{15}$ & $4 \times 10^{18}$ \\
white dwarf star   & $1 ~ \times ~ 10^{17}$ & $4 \times 10^{18}$ \\
neutron star       & $1 ~ \times ~ 10^{20}$ & $6 \times 10^{18}$ \\
\hline
\end{tabular}
\end{center}
\end{table}

To judge the effectiveness of accretion as an energy source, Table 2 
compares the gravitational potential energy, $\epsilon_g = G \mstar/\rstar$,
of various 1~\msun~stars with the nuclear energy generation rate of the 
dominant fusion source, $\epsilon_{nuc}$.  If a star can accrete {\it and burn} 
material from a disk, then the ratio of the accretion luminosity to the
nuclear luminosity is simply the ratio of these two quantities,
$r \equiv L_{acc}/L_{nuc}$ = $\epsilon_g/\epsilon_{nuc}$.  This expression
ignores the difference in accretion and nuclear time scales, but it
provides a simple guide to the importance of accretion as a long-term
energy source.  Table 2 shows that accretion is rarely the major energy
source for main sequence stars and white dwarfs; nuclear energy will
always overwhelm accretion by factors of 40--2000 on long time scales.
Accretion is an excellent energy source for pre-main sequence stars
and neutron stars.  It is roughly 10 times more effective than 
deuterium burning in pre-main sequence stars and nearly 20 times
more productive than complete hydrogen burning in neutron stars.

\subsection{Turbulent Viscosity and Disk Timescales}

The source of the frictional heating in accretion disks remains
controversial.  Ordinary molecular viscosity is too small to 
generate mass motion on a reasonable time scale, $\sim$ days
for disks in interacting binary systems  and $\sim$ years to 
decades for disks in pre-main sequence stars.  
The large shear between adjacent disk annuli suggests that disks
might be unstable to turbulent motions,
which has led to many turbulent viscosity mechanisms.  
Convective eddies, gravitational instabilities, 
internal shocks, magnetic stresses, sound waves, spiral density waves, 
and tidal forces have all been popular turbulence mechanisms 
in the past three decades
(see, for example, \cite{lyn74}, \cite{sha73}, \cite{cab87}, 
\cite{lin87}, \cite{vis89}, \cite{bal91}, \cite{haw91}, \cite{tou92}, 
\cite{sto96}, \cite{cab96}, \cite{vis96}, \cite{roz96}, \cite{reg97},
\cite{arm98}).

Recent work has shown that magnetic stresses in a differentially
rotating disk inevitably lead to turbulence (\cite{bal96}, 
\cite{bal98}, \cite{lin95}).
The growth time and effectiveness of these magnetohydrodynamic
mechanisms make them the current leading candidate for viscosity 
in most applications.
How this turbulence leads to significant mass motion in a real
accretion disk remains an unsolved problem.

Shakura \& Sunyaev side-stepped the basic uncertainties of viscosity 
mechanisms when they developed the popular ``$\alpha$-disk'' model
(\cite{sha73}; see also \cite{von48}).  
In this approach, the frictional heating in the disk 
is due to a turbulent viscosity, $\nu = \alpha c_s h$, where $c_s$ is the 
sound speed, $h$ is the local scale height of the gas, and $\alpha$ 
is a dimensionless constant. This concept is similar to the mixing 
length theory of convection, with $\alpha$ serving the role of the
mixing length.   In most applications, $\alpha \simless$ 1--10;
$\alpha$ needs to exceed $\sim 10^{-4}$ to allow material to move 
inwards on a reasonable time scale.

This viscosity definition orders the important time scales for
the disk (\cite{lyn74}).  The shortest disk time scale is the 
dynamical (orbital) time scale, which increases radially outward:

\begin{equation}
\tau_D \approx {\rm 0.1~day}~ \left ( \frac{R}{1~\rsun} \right )^{3/2} \left ( \frac{1~\msun}{\mstar} \right )^{1/2}
\end{equation}

\noindent
The thermal time scale measures the rate that energy diffuses through 
the disk,

\begin{equation}
\tau_T \approx {\rm 1~day}~ \left ( \frac{R}{1~\rsun} \right )^{11/8} ~ .
\end{equation}

\noindent
The thermal time scale is intermediate between the dynamical time
scale and the viscous time scale, which measures the rate matter
diffuses through the disk,

\epsfxsize=5.5in 
\hskip 12ex 
\epsffile{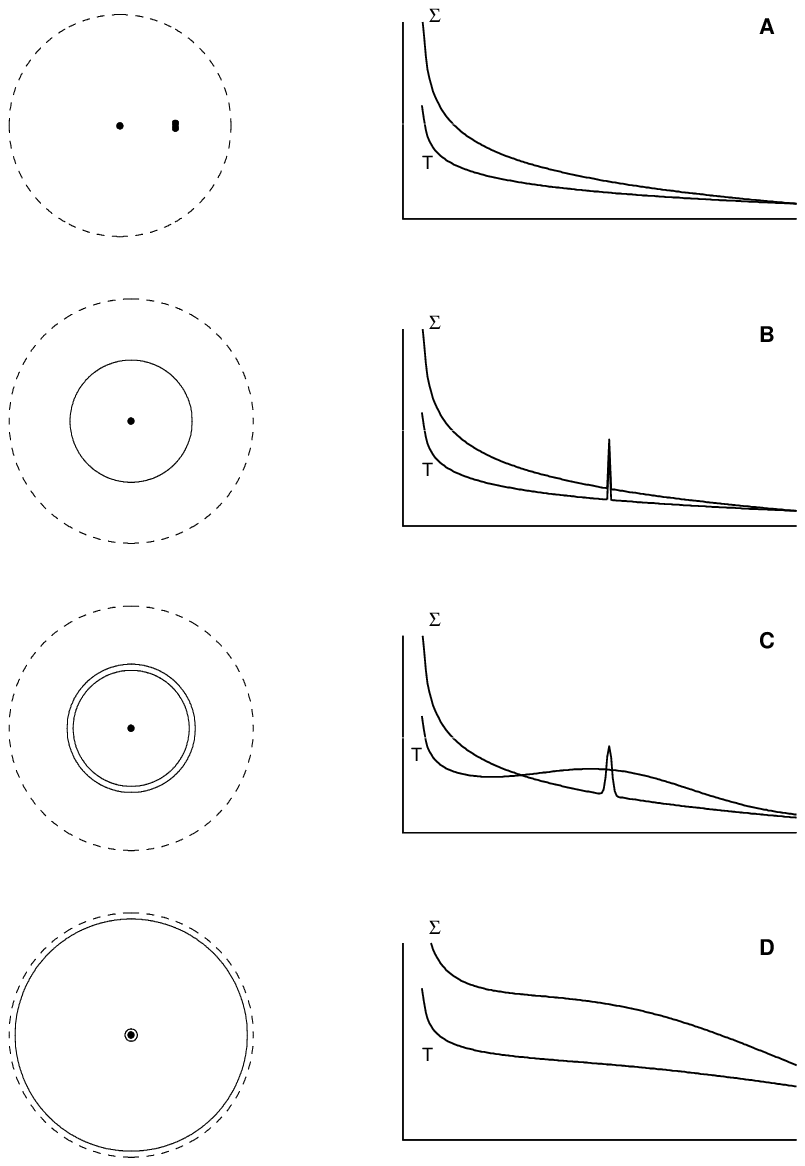}
 
\footnotesize
\footspace

\vskip -4ex
\noindent{Figure 5 - Schematic evolution of a point-like mass enhancement, 
a blob, in a viscous accretion disk.  A central star lies at the center of 
the disk with an outer edge indicated by the dashed circle.
The disk initially has a smooth
radial decrease in the effective temperature, T, and the surface density,
$\Sigma$ (panel A).  Differential rotation between adjacent annuli smooths
the blob into a ring on the orbital time scale, $\tau_D$, which produces
$\delta$-function increases in T and $\Sigma$ (panel B).  The thermal energy 
in the blob moves inwards and outwards on the thermal time scale, $\tau_T$, 
which produces a gaussian-like perturbation in the temperature (panel C).  
The mass motion on this time scale is small; $\Sigma$ changes little on 
the thermal time scale.  The mass perturbation spreads out through the
disk on the viscous time scale, $\tau_V$, as indicated in panel D.}

\vskip 4ex
\normalsize
\singlespace

\begin{equation}
\tau_V \approx \frac{\rm 40~day}{\alpha} ~ \left ( \frac{R}{1~\rsun} \right )^{5/4} 
\end{equation}

Figure 5 illustrates the evolution of the disk on these time scales.
The top panel shows a blob of material superimposed on the smooth density
and temperature structure of a steady disk. In this example, the blob has a 
mass, $\delta M$, larger than the mass of an annulus, and a thermal energy
content, $\delta E$, larger than the local thermal energy of an annulus.
Disk rotation broadens the blob into a narrow ring in one or two rotation 
periods, which produces a narrow spike in the radial distributions of 
surface density and temperature.  The thermal energy of this ring moves
inwards and outwards on the thermal time scale. The temperature
distribution broadens considerably, but the surface density hardly
changes.  The mass in the ring finally moves inwards and outwards on 
the viscous time scale. This evolution produces an extra increase in
the temperature due to viscous dissipation. 

Table 3 compares numerical values for the three disk time scales.
These results assume \mstar~= 1 \msun~ and $\alpha = 10^{-1}$.
For most interacting binaries, the disk radius is $\sim$ 0.1--10 
\rsun; the time scales range from a fraction of a day for $\tau_D$ 
up to several tens of days for $\tau_V$.  The time scales increase 
dramatically for the larger disks in pre-main sequence stars and 
active galactic nuclei.  The viscous time scale at the edge of a 
typical solar system is comparable to the disk lifetime.

\begin{table}[htb]
\begin{center}
\caption{Disk time scales}
\begin{tabular}{lcccccc}
\hline 
Disk radius & h/r & T (K) & $\tau_D$ & $\tau_T$ & $\tau_V$ \\
\hline
0.1 \rsun & 0.02 & $2 \times 10^4$ & 0.004 d & ~~0.04 d & ~20 d \\
1.0       & 0.02 & $5 \times 10^3$ & 0.120 d & ~~1.00 d & 400 d \\
\\
1 AU     & 0.04 & 250 & ~~~1 yr & ~~~~~4 yr & $1 \times 10^4$ yr \\
10       & 0.06 & ~60 & ~~32 yr & ~~~105 yr & $1 \times 10^5$ yr  \\
100      & 0.08 & ~25 & 1000 yr & ~~2500 yr & $2 \times 10^6$ yr  \\
\hline
\end{tabular}
\end{center}
\end{table}

\subsection{Disk Energy Distributions}

Figure 6 shows the broadband spectral energy distribution of a steady 
accretion disk with a boundary layer.  The model assumes that the disk 
is optically thick and that each annulus radiates as a blackbody 
(\cite{lyn74}, \cite{bat74}, \cite{tyl81}, \cite{ruc85}).  There are 
two main features in this spectrum:
(i) radiation from the boundary layer at short wavelengths, and
(ii) disk radiation at longer wavelengths.  If the disk has a large
ratio of outer radius to inner radius, $R_{out}/R_{in} \gg 1$, the
disk spectrum follows

\begin{equation}
\lambda F_{\lambda} \propto \lambda^{-4/3} ~~ \hspace{20mm} ~~ \lambda_{in} < \lambda < \lambda_{out}
\end{equation}

\noindent
where $\lambda_{in} T_d(R_{in}) \approx \lambda_{out} T_d(R_{out}) \approx$ 
0.36 (\cite{lyn74}).  The disk spectrum follows the Wien tail of a hot
blackbody, $T_d(R_{in})$, at short wavelengths and the Rayleigh-Jeans 
tail of a cool blackbody, $T_d(R_{out})$, at long wavelengths.  

Although Figure 6 assumes a blackbody disk, the general features of the
model change little if the disk radiates as some type of stellar
atmosphere.  The most important modifications for pre-main sequence
stars involve irradiation from the central star and the infalling
envelope, as described by Beckwith in this volume.

\epsfxsize=5.in 
\hskip 12ex 
\epsffile{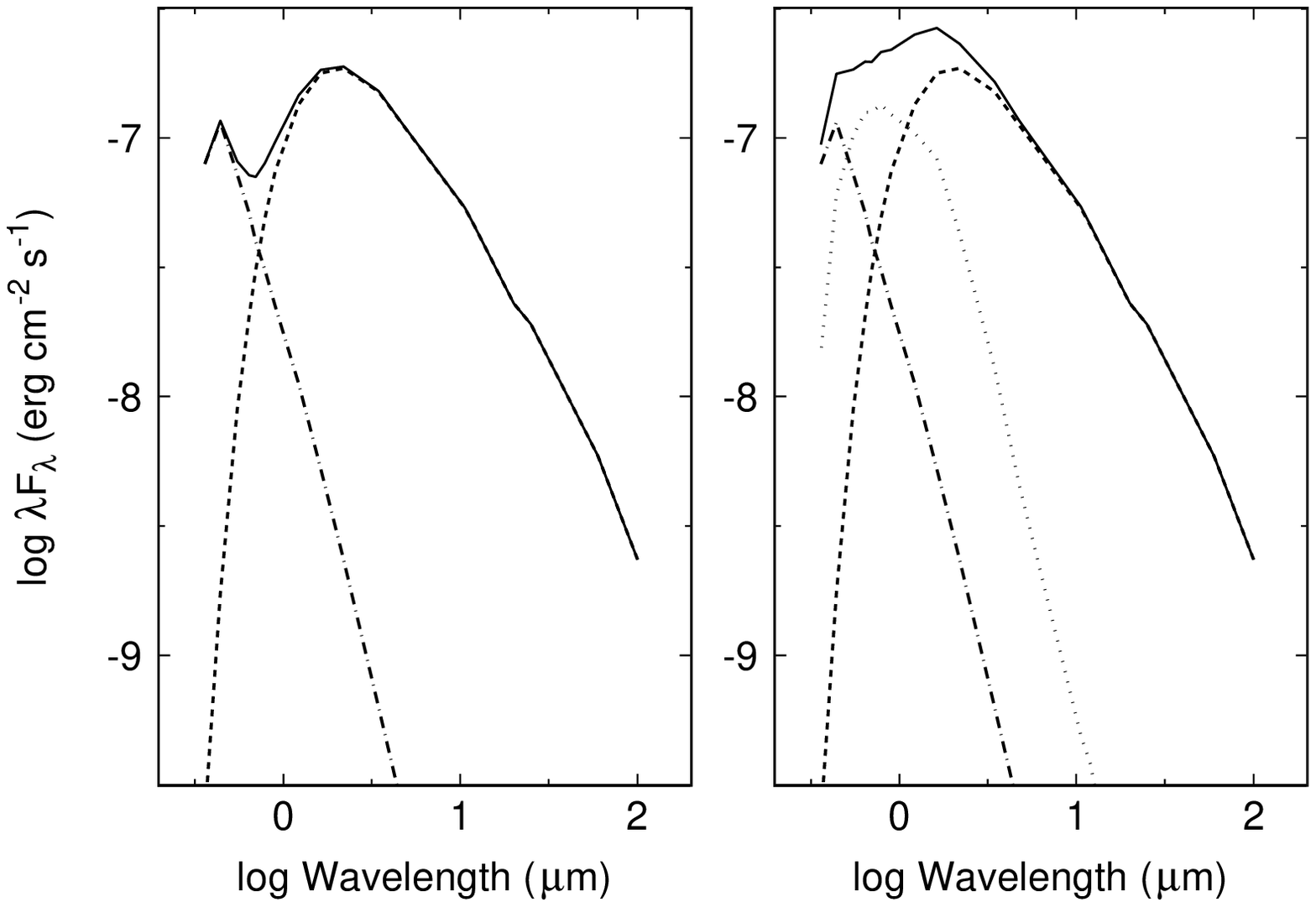}
 
\footnotesize
\footspace

\vskip -7ex
\noindent{Figure 6 - Energy distributions for accretion disks.
The left panel shows the energy distribution of a disk (dashed line),
a boundary layer (dot-dashed line), and the total energy output
(solid line) for \mdot~= $2~\times~10^{-7}~\msunyr$ onto a solar-type star.  
The right panel adds the energy distribution (dotted line) 
for a star with $L_{\star} \approx L_d$.

\vskip 3ex
\normalsize
\singlespace

Lynden-Bell \& Pringle first applied these concepts to disks in
interacting binaries and pre-main sequence stars.  They added radiation
from the central star to the spectrum, as indicated in the right panel 
of Figure 6.  This model suggests that identifying disk radiation can 
be difficult when the stellar luminosity exceeds the accretion luminosity.
Table 4 compares temperatures and luminosities of pre-main sequence stars and 
steady disks for two accretions rates, $\mdot = 10^{-6}$ and $10^{-4}~\msunyr$.
At low masses, the contrast between the disk and star is considerable 
unless the accretion rate is $\simless 10^{-8}~\msunyr$.  This contrast 
decreases for more massive stars.  At 10 \msun, the star is hotter 
and as luminous as the disk, because massive young stars burn hydrogen 
before reaching the main sequence.  Table 3 shows that accretion cannot 
be a dominant energy source for main sequence stars; the difficulty of
seeing an accretion disk against the background of a young O or B star
therefore is not surprising.

\begin{table}[htb]
\begin{center}
\caption{Accretion and stellar luminosities in pre-main sequence stars}
\begin{tabular}{lccccccc}
\hline 
 & & & & \multicolumn{2}{c}{$10^{-6}~\msunyr$} & \multicolumn{2}{c}{$10^{-4}~\msunyr$} \\
\mstar (\msun) & \tstar (K) & \lstar (\lsun) & & $T_{max}$ (K) & $L_{acc}$ (\lsun) & $T_{max}$ (K) & $L_{acc}$ (\lsun) \\
\hline
0.1 & 3100 & 0.1 & & 1850 & 3 & 5850 & 275 \\ 
0.4 & 3600 & 0.5 & & 1825 & 7 & 5800 & 675 \\
1.0 & 4250 & 7.0 & & 1800 & 12 & 5700 & 1200 \\
3.0 & 8650 & 80 & & 1700 & 25 & 5350 & 2350 \\
6.0 & 17000 & 1000 & & 2100 & 50 & 6650 & 5000 \\
10.0 & 21000 & 5000 & & 1850 & 60 & 5850 & 6000 \\
20.0 & 29000 & 45000 & & 1600 & 75 & 5000 & 7450 \\
\hline
\end{tabular}
\end{center}
\end{table}

\section{Unstable Accretion Disks}

Despite the elegant simplicity of the steady-state model, 
all disks vary their energy output.  These fluctuations range
from small, 10\%--20\%, amplitude flickering on the local 
dynamical time scale (\cite{bru94}) to large-scale eruptions,
$\sim$ 3--5 mag or more, that can last for several times the local 
viscous time scale (Figure 1).  In well-studied dwarf novae,
the outbursts occur in cyclical patterns with quasi-periods
of 10 or more days (\cite{can98}, \cite{opp98}).  

Osaki and Pringle \etal first noted the possibility for thermal
instabilities in accretion disks (\cite{pri73}, \cite{osa74}).  
In the standard picture, the structure of a steady disk is 
derived from balancing heat generated by viscosity with 
radiative cooling.  However, the viscous energy generation 
and radiative losses are set by local disk parameters; the
input \mdot~is an external parameter.  Cooling can balance heating 
only if the local physical quantities can adjust to the input $\mdot$.
If the local physical variables cannot adjust, cooling cannot 
balance heating and a thermal instability is likely.

Figure 7 displays a simple illustration of a thermal instability 
in a single disk annulus.  The solid curves indicate loci of
stable disks, where cooling precisely balances heating.
To the right of these lines, heating exceeds cooling because 
viscous energy generation -- set by the local surface density
$\Sigma$ through $\mdot = \nu \Sigma$ -- exceeds the radiative 
losses set by the effective temperature, $T_d$.  Cooling exceeds 
heating to the left of the stability lines.

To illustrate the evolution of the instability, assume the
annulus has ($\Sigma$, $T_d$) at A$^{\prime}$ and an input accretion rate, 
$\mdot_1$. This input accretion rate exceeds the stable accretion 
rate at A$^{\prime}$, $\mdot_s = \mdot_{A^\prime}$.  The annulus 
evolves up the 
solid line on the viscous time scale to find a solution where 
$\mdot_s = \mdot_1$.  Before reaching this point, the annulus
arrives at $\mdot_B$, the last stable solution for this opacity source.
At B, the annulus needs a larger surface density to accommodate the 
larger $\mdot$, and a larger $T_d$ to radiate away the extra heat 
generated by the larger $\mdot$.  To do so, the annulus evolves to
C; it makes this transition at constant $\Sigma$ because the
thermal time scale is shorter than the viscous time scale.
Once at C, the annulus radiates more energy than generated by
viscous dissipation ($\mdot_s = \mdot_C > \mdot_1$) and evolves 
down the solid line towards a solution where $\mdot_s = \mdot_1$.  
The annulus moves towards lower $\Sigma$ on the viscous time scale 
until it reaches point D, the last stable solution on this branch of 
the equilibrium curve.  At D, the annulus drops to A to
try to find the smaller $T_d$ needed to accommodate a smaller 
$\mdot$.  Once at A, the annulus retraces the steps around this 
`limit-cycle' curve as long as $\mdot = \mdot_1$.

\epsfxsize=5.4in
\epsffile{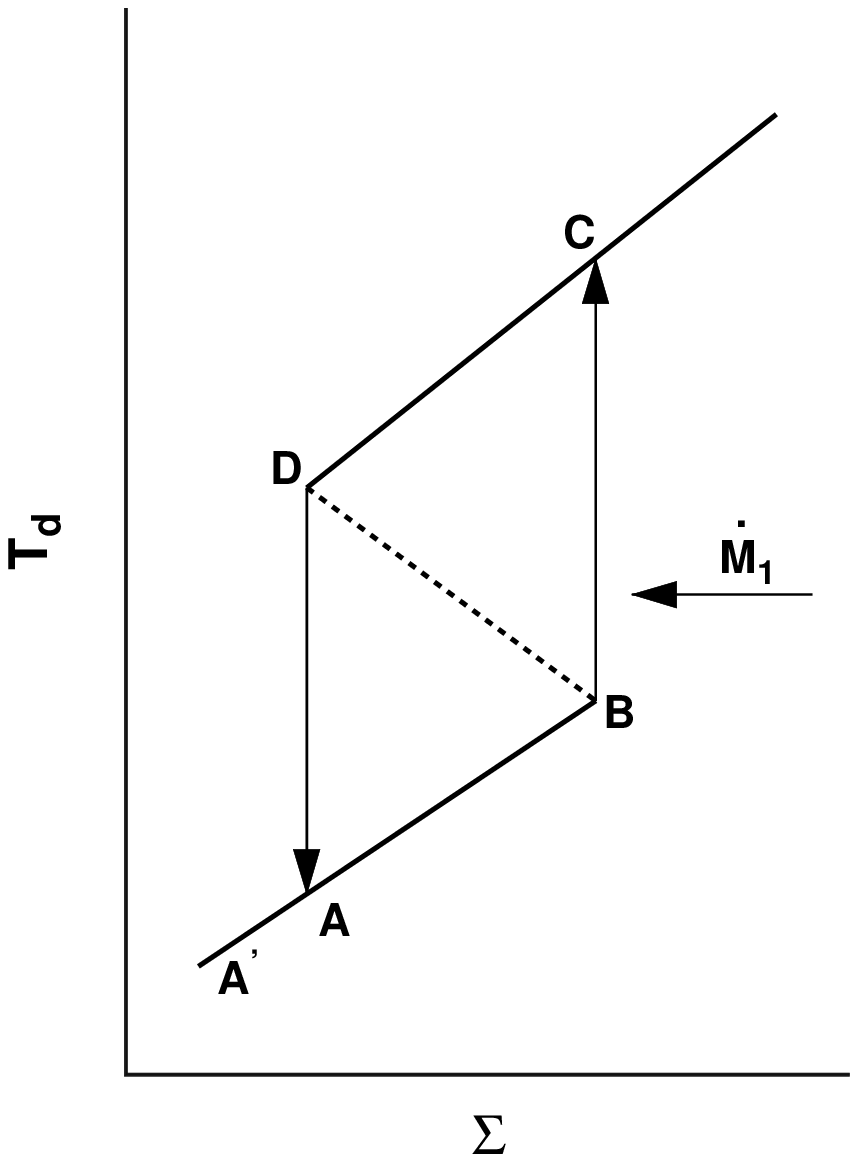}

\footnotesize
\footspace

\vskip -2ex
\noindent{Figure 7 - Outline of a disk instability in a single annulus.
The disk evolves along the solid lines ABCD for an input $\mdot$
(horizontal arrow). Heavy solid lines correspond to stable equilibrium 
solutions; a dashed line indicates the locus of unstable equilibria.}

\vskip 4ex
\normalsize
\singlespace

The evolution of an instability in a complete accretion disk
follows the simple illustration.  In a real disk, most annuli 
are close to B when a single annulus makes the transition from 
B to C.  The increased temperature of a single annulus transfers 
heat to neighboring annuli on the thermal time scale; these annuli 
then jump to the `high state' and propagate the eruption to their 
neighbors.  Figure 8 shows this evolution for five snapshots in 
the evolution of a dwarf nova accretion disk.  Time increases upwards 
in the Figure. The bottom panel shows the temperature distribution 
of the disk at the onset of the eruption.  The inner edge of the 
disk jumps to the high state first; annuli at larger radii 
progressively follow until all of the disk resides in the 
high state in the top panel. The disk maintains this state for 
several days, and then retraces its path back to the low state.

\epsfxsize=7.5in
\epsffile{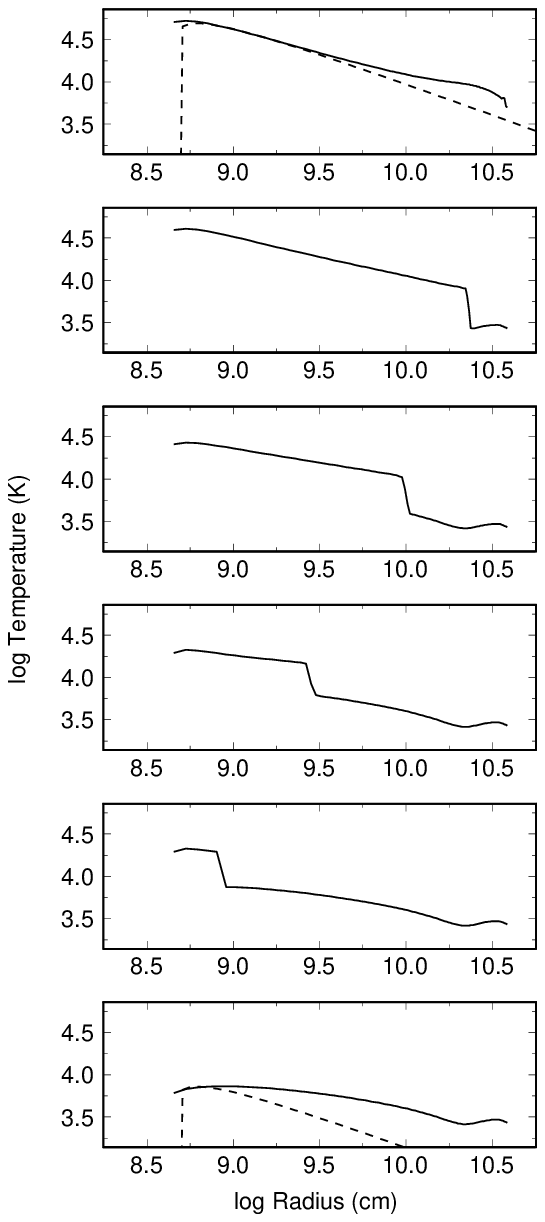}

\footnotesize
\footspace

\noindent{Figure 8 - Evolution of a disk instability in a dwarf
nova disk.  The bottom panel compares the radial temperature 
distribution of a time-dependent disk (solid line) with the
temperature distribution of a steady disk (dashed line). The 
eruption begins in the next panel with a sharp increase in the
temperature at the inner edge of the disk.  This increase creates
a `wave' of increased temperature that propagates radially outwards 
in the disk as time moves forward (up in the figure). In the top
panel, the entire disk has reached the hot state; the actual 
temperature distribution (solid line) is then close to the steady-state
temperature distribution (dashed line)}

\vskip 4ex
\normalsize
\singlespace

One feature of the disk instability picture is that the disk is
never in a steady-state.  The dashed lines in Figure 8 plot
steady-state temperature distributions normalized to the
temperature at the inner edge of the disk.  The actual temperature 
gradient is much flatter than the steady-state gradient in 
the low state (bottom panel).  The $\mdot$ through the disk 
increases radially outwards and is below the input $\mdot$ 
at each point in the disk.  The disk approaches but never reaches 
the steady-state gradient in the high state (top panel).

Actual disk temperature distributions have been derived from 
eclipse light curves similar to those shown in Figure 1 for CI Cyg.
K. Horne \& collaborators (\cite{ho85a}, \cite{ho85b}, \cite{mar88},
\cite{bap98}) have used maximum entropy techniques to recover $T_d(R)$
from multi-wavelength light curves and spectroscopic data assuming 
(i) the disk is azimuthally symmetric and (ii) the brightness
temperature is close to the local blackbody temperature.
Their results indicate that quiescent disks are rarely close to 
steady-state, with temperature distributions usually much flatter 
than $T_d(R) \propto R^{-3/4}$.
Systems in eruption -- dwarf novae and symbiotic stars at
maximum light -- more closely resemble, but never achieve, 
the steady-state temperature distribution with $q$ = 3/4.
These results generally agree with the model predictions in Figure 8.

\section{Disk Eruptions in Pre-Main Sequence Stars}

Most pre-main sequence stars vary in brightness.  Irregular
brightness variations of $\simless$ 1 mag are a defining
feature of T Tauri stars and many Herbig AeBe stars 
(\cite{joy45}, \cite{her60}).  Recent studies show that many
of these variations are due to dark spots rotating with the
stellar photosphere (\cite{hrb87}, \cite{bou89}, \cite{bou93}, \cite{hrb94}) 
or bright spots produced at the base of a magnetic accretion column
(\cite{bou89}, \cite{ken94}).  Small eruptions of 1--3 mag 
lasting several years occur in the EXors, a poorly-studied class 
which includes EX Lup and DR Tau (\cite{her89}).  More
spectacular 3--6 mag eruptions occur in the FU Orionis variables,
also known as FUors (\cite{her77}, \cite{har96}).

G. Herbig first associated the eruptions of FUors with 
pre-main sequence stars (\cite{her66}, \cite{her77};
see also \cite{amb54}, \cite{kho59}).
FU Ori, which lies at the apex of a fan-shaped nebula 
within the dark cloud B35, brightened by a factor of 
over one hundred in 100--200 days (Figure 9; \cite{wac39},
\cite{wac54}, \cite{her66}, \cite{her77}, \cite{ibr93}, 
\cite{she95}, \cite{ibr97}).
Thirty years later, Welin discovered the 5 mag brightening
of V1057 Cyg within an eccentric ring of reflection nebulosity
(Figure 9; \cite{wel71}, \cite{her77}, \cite{ken91a}, \cite{kol97}).  
Herbig later noted the similarity between the optical 
spectra of these two stars with spectra of V1515 Cyg, 
a faint variable star embedded in arc-shaped nebulosity 
(\cite{her60}).  He collected archival photographic photometry 
and identified a slow rise from $m_{pg} \approx$ 15.5 in the 
late 1940s to $m_{pg} \approx$ 13.5 in the late 1970s 
(\cite{her77}).  This brightness increase continued until 1980,
when the star experienced a dramatic decline and slow recovery 
(\cite{kol83}, \cite{ken91b}).

Herbig's demonstration that FU Ori -- and by analogy other FUors --
is a pre-main sequence star is straightforward.  FUors are clearly 
associated with dark clouds, having radial velocities 
indistinguishable from the cloud velocity.  The optical spectrum,
including the high lithium abundance, has similarities with spectra 
of T Tauri stars.  The event statistics are plausible for pre-main
sequence stars.  Finally, the rise in brightness is a real luminosity 
increase that is {\it not} a nova outburst, the main alternative.

\vskip 2ex
\epsfxsize=7.0in
\epsffile{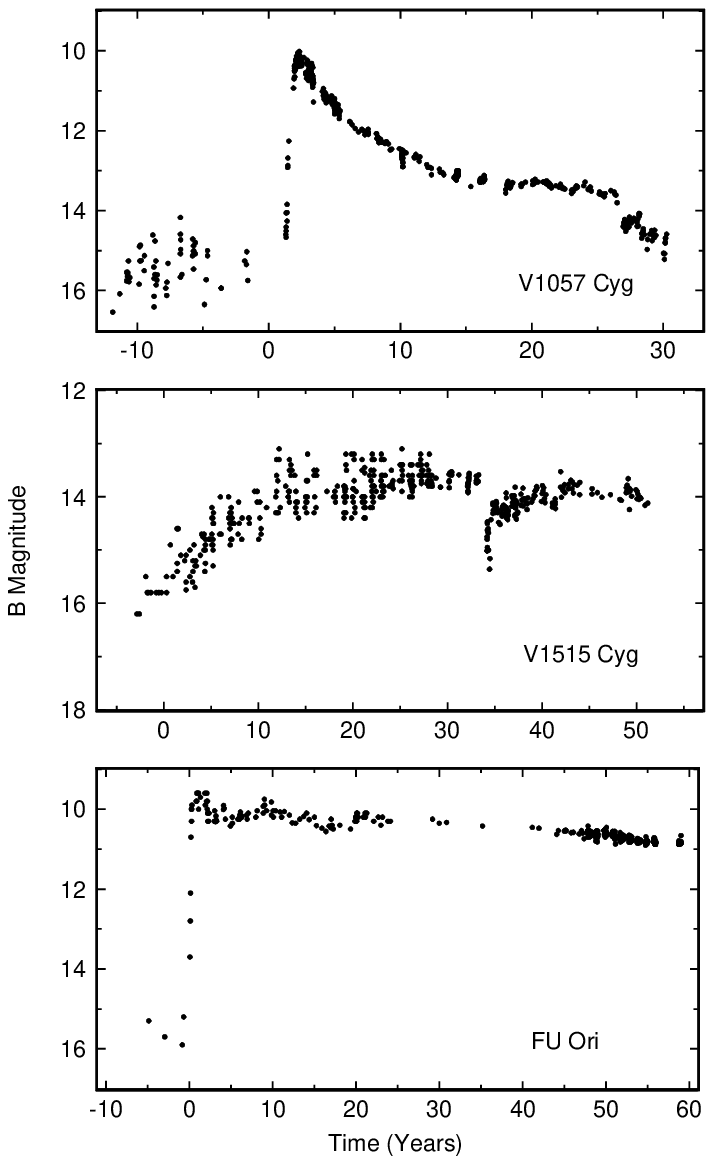}

\footnotesize
\footspace

\vskip -6ex
\centerline{Figure 9 - Blue light curves for three FU Ori variables.}

\vskip 4ex
\normalsize
\singlespace

Many new FUors have been discovered since V1515 Cyg. The catalogued 
number now stands at 11 (\cite{har96}, Table 5). 
Most have been observed to rise 3--5 mag in optical or near-IR brightness
in less than one year, as indicated by the `Y' in the outburst (OB) column
of Table 5. V1515 Cyg is the 
only known example to require a decade to rise to visual
maximum, but the historical light curves for some systems
are poorly documented.
A few objects have been called FUors based on common spectroscopic
properties, which are described more completely below.  
Most recent FUors are more intimately associated with the densest
dark clouds than the first members of the class, which 
suggests that many eruptions might have been missed.

There are also 3 candidate FUors listed as a second group
of objects in Table 5.  These objects have some properties in 
common with FUors, but more data are needed to see if they have
the other characteristics as well (\cite{str93}, \cite{san98}).
A few other objects have one or two properties of known FUors,
such as an outburst or unusual light curve (e.g., 
\cite{eis91}, \cite{hod96},
\cite{alv97}, (\cite{she97}, \cite{yun97}).
These are more probably EXors or Ae/Be stars than FUors.

\subsection{Basic Properties of FU Orionis Objects}

FUors share a distinctive set of morphological,
photometric, and spectroscopic characteristics (Table 5). 
Three are known binary stars, L1551 IRS5, Z CMa, and 
RNO 1B/1C (\cite{kor91}, \cite{ken93b}, \cite{bar94}, 
\cite{thi95}, \cite{rod98}).
Most have delicate fan-shaped or comma-shaped reflection 
nebulae (RN) on optical and near-IR images (\cite{goo87}, \cite{rei91},
\cite{nak95}, \cite{luc96}).
Optical jets and HH objects are common (\cite{rei91}, \cite{dav94},
\cite{har96}).
Most appear associated with large-scale molecular outflows 
(\cite{yam92},
\cite{fri93}, \cite{eva94}, \cite{mcm95}, \cite{yan95}, 
\cite{fri98}, \cite{lop98}, and references therein).  
These features -- together with the broad, 
blue-shifted Na~I and H~I absorption features  described below -- 
demonstrate that FUors drive powerful winds that interact with 
the surrounding medium (\cite{bas85}, \cite{cro87}, \cite{whi93}).

\begin{table}[htb]
\begin{center}
\caption{Selected properties of FU Orionis objects}
\begin{tabular}{lcccccccc}
\hline 
Object & OB & Opt ST & IR ST & RN & HH/Jet & Wind & Outflow & Radio \\
\hline
L1551 IRS 5 & ? & K & M & Y & Y & ? & Y & Y \\
FU Ori      & Y & F--G & M & Y & N & Y & N & N \\
Z CMa       & ? & F--G & M & Y & Y & Y & Y & Y \\
BBW 76      & @ & G & M & Y & ? & Y & N & ? \\
V346 Nor    & Y & ? & C & Y & Y & ? & Y & ? \\
Par 21      & ? & F & M & Y & Y & Y & ? & ? \\
V1515 Cyg   & Y & G & M & Y & N & Y & Y & N \\
V1057 Cyg   & Y & G & M & Y & N & Y & Y & Y \\
V1735 Cyg   & Y & ? & M & Y & N & ? & Y & Y \\
RNO 1B/1C   & Y & G & M & Y & N & ? & Y & ? \\
\\
IC 430      & ? & G & ? & Y & Y & Y & N & Y \\
PP 13S      & Y & ? & M & ? & ? & ? & Y & ? \\
Re 50       & ? & G & ? & Y & Y & Y & Y & Y \\
\hline
\end{tabular}
\end{center}
\end{table}

FUors also have very unusual spectroscopic characteristics.
The optically visible sources have F--G, and sometimes K-type,
giant or supergiant spectra (\cite{her77}; Table 5, see also
\cite{har88}, \cite{re97a}, \cite{re97d})). 
The optical reflection nebulae of several embedded FUors 
also show G-type absorption features (\cite{sto88}, 
\cite{sta92}).  All FUors but Z CMa and V346 Nor have very deep 
CO absorption bands on near-IR spectra at 1.6 $\mu$m and 2.3 $\mu$m
(\cite{mou78}, \cite{eli78}, \cite{car87}, \cite{sat92},
\cite{ken93a}, \cite{bis97}).
These features resemble the CO absorption bands observed in 
red giants and are much stronger than those observed in any 
other pre-main sequence star (\cite{har87a}, \cite{har87b},
\cite{gre97}).  CO absorption in Z CMa is weakened by dust emission 
from an embedded companion; V346 Nor has weak CO emission instead
of CO absorption (\cite{kor91}, \cite{teo97}; \cite{re97d}).
Many FUors also display strong water absorption features, which 
strengthens the evidence for a $\sim$ 2000 K photosphere
(\cite{cal91}, \cite{sat92}, \cite{cal93}).

All FUors show large {\it excesses} of radiation over normal 
G supergiants at both ultraviolet and infrared wavelengths.
The near-IR excess is clearly photospheric in origin, 
because the CO and H$_2$O absorption features are so intense.
The UV excesses in Z CMa and FU Ori appear associated with
an A- or F-type photosphere that is hotter than the G-type
photosphere observed at longer wavelengths (\cite{ke89b}).
In addition to significant far-IR and submm emission
(\cite{wei89}, \cite{wei91}, \cite{ken91a}),  many FUors are
strong radio continuum sources at cm wavelengths (\cite{coh82},
\cite{rod90}, \cite{rod92}, \cite{loo97}, \cite{rod98}).
This emission is not photospheric in some FUors and
may be produced in the outflow or the jet.

Herbig first noted broad absorption lines on optical spectra 
of several FUors (\cite{her77}).  Hartmann \& Kenyon later
confirmed large rotational velocities of $v$ sin $i$ $\approx$ 
15--60 \kms~for V1057 Cyg, V1515 Cyg, and FU Ori (\cite{har85}).
This property is now characteristic of the class; all FUors have 
broad optical or infrared absorption lines or both (\cite{har88},
\cite{sta91}, \cite{sta92}, \cite{ken93a}).
The rotational velocity further depends on wavelength.
The near-IR CO lines in FU Ori 
and V1057 Cyg have significantly smaller rotational velocities 
than the optical lines (\cite{har87a}, \cite{har87b}).  The 
optical rotational velocity smoothly increases with decreasing 
wavelength in Z CMa and V1057 Cyg (\cite{wel91}, \cite{wel92}).
In FU Ori itself, a powerful wind masks weak absorption lines and
makes it difficult to detect any variation of rotational velocity
with wavelength should one exist.

Many FUors display {\it doubled} absorption lines on optical 
and near-IR spectra (\cite{har85}, \cite{har87a}, \cite{har87b},
\cite{har88}, \cite{sta91}, \cite{sta92}).
The two absorption components in V1057 Cyg are separated by 
30--40 \kms; these features have much larger separations in 
Z CMa and FU Ori.  The long-term stability of the doubled
absorption lines indicates that the lines are not produced by
two stellar components in a binary system (\cite{ken88}).

Most FUors also show strong evidence for mass loss, in addition
to the HH objects and CO outflows described above.
Practically every FUor displays deep, blueshifted absorption
components on H$\alpha$ and Na I D (\cite{bas85}, \cite{cro87}, 
\cite{eis90}, 
\cite{wel92}, \cite{har95}); weak blueshifted CO absorption 
might be present in FU Ori (\cite{har87b}).  The optical line 
profiles can also change on month to year time scales 
(\cite{bas85}, \cite{cro87}, \cite{wel92}).  The near-IR
line profiles may also change (\cite{bis97}).
Both V1057 Cyg and V1515 Cyg have very pronounced line 
profile changes and dips in their optical light curves.
These fluctuations suggest dust formation in a variable wind,
but the data are not very extensive (\cite{ken91b}).

Finally, FUor eruptions must be repetitive (\cite{her77}, 
\cite{har85}).  The event statistics of known FUors suggest
that a young star must undergo 10--20 FUor eruptions before 
reaching the main sequence.  This estimate may be a lower 
limit, because some outbursts have certainly
been missed.  Reipurth's proposal that FUor eruptions produce
HH objects suggests a recurrence time scale of $\sim$ 1000 yr,
based on the identification of multiple bowshocks (ejection
events) with dynamical separations of 500--2000 yr
(\cite{rei85}, \cite{rei89}, \cite{hrt90}, \cite{rei92},
\cite{bac94}).
The discovery of giant HH flows with dynamical time scales
of $10^4$--$10^5$ yr (\cite{re97c}) allows 10--100 eruptions
if eruptions actually power the outflow (see below).

\subsection{FU Orionis Objects as Accretion Disks}

The observations described above place severe constraints 
on possible FUor outburst mechanisms.  With a luminosity
of a few hundred \lsun~(Table 6), a typical FUor emits
$10^{45}$ to $10^{46}$ erg during the course of a 10--100 yr
eruption.  Outbursts also recur every $10^3$--$10^5$ yr.  
This process produces an object resembling a rapidly rotating 
F-G supergiant in the optical and a more slowly rotating
M giant in the infrared.
It must be unique to young stars, because we have not observed 
a FUor event in an older star system.

Accretion is the most plausible energy source for FUor eruptions
(\cite{har85}, \cite{lin85}).  By analogy with dwarf novae,
FUor eruptions are the high state of a thermal instability in
a disk surrounding a pre-main sequence star.  This instability
occurs if the accretion rate from the molecular cloud core into
the disk lies between the stable accretion rates in the low state
and the high state (Figure 7).  The system cycles between these 
two states as long as the surrounding cloud can replenish the disk 
between outbursts.

\begin{table}[htb]
\begin{center}
\caption{Disk Properties of FU Orionis Objects}
\begin{tabular}{lccc}
\hline 
Object & $L_{bol}$ (\lsun) & \mstar \mdot ($10^{-5}~\mssunyr$) & \rstar (\rsun) \\
\hline
L1551 IRS 5 &  30 & 0.1 & 0.4 \\
FU Ori      & 220 & 1.7 & 1.2 \\
Z CMa       & 400 & 4.0 & 1.6 \\
BBW 76      & 200 & 1.4 & 1.1 \\
V346 Nor    & 160 & 1.0 & 1.0 \\
Par 21      & 200 & 1.4 & 1.1 \\
V1515 Cyg   & 135 & 0.8 & 0.9 \\
V1057 Cyg   & 400 & 4.0 & 1.6 \\
V1735 Cyg   & 200 & 1.3 & 1.1 \\
RNO 1B/1C   & 750 & 3.5 & 1.5 \\
\hline
\end{tabular}
\end{center}
\end{table}

Table 6 lists the disk properties needed to power observed FUor
luminosities from accretion.  To make these estimates, I assume 
a steady disk with $L_{disk} = 157~\lsun (\mstar \mdot / \rstar) $
and $T_{max} \approx$ 6500 K $(\mstar \mdot / \rstar^3)^{1/4} $.
Combining these and setting $T_{max}$ = 6500 K for an F-G supergiant
star yields

\begin{equation}
\rstar \approx \left ( \frac{L_{disk}}{157~\lsun} \right )^{1/2}
\end{equation} 

\noindent
and
 
\begin{equation}
\mstar \mdot \approx \left ( \frac{L_{disk}}{157~\lsun} \right )^{3/2}
\end{equation}

\noindent
These expressions produce the results in Table 6 for known FUors,
assuming a typical inclination angle, cos $i$ = 1/2.
The median inner radius for the FUor sample is
\rstar $\approx$ 1.1--1.2 \rsun; the median mass
accretion rate is $\mstar \mdot \sim$ 1--2 $\times~10^{-5}~\mssunyr$.
For a mass-radius relation for pre-main sequence stars with ages 
of $10^5$ yr, $\mstar \approx$ 0.2--0.3 \msun~and
$\mdot \approx$ 0.5--1 $\times~10^{-4}~\msunyr$.

This very simple analysis indicates that a young star must accrete 
material at $\sim 10^{-4}~\msunyr$ during an eruption to power the 
observed luminosity.  For an adopted outburst duration of 100 yr, 
this accretion rate results in a total accreted mass of $\sim$ 
0.01 \msun~per eruption.  
The total mass accreted during a FUor eruption must be
replenished during a low state lasting $\sim$ $10^3$--$10^5$ yr,
which results in an infall rate of $\sim$ $10^{-5}$--$10^{-7}~\msunyr$.
This range is close to the infall rates envisioned -- and in some cases
observed -- for typical cloud cores (\cite{ada87}, \cite{oha96},
\cite{mar97}).

\subsection{Observational Tests of Disk Models}

Steady accretion disk models successfully explain many observations 
of FUors (\cite{har85}, \cite{ken88}, \cite{har96}).  
Theoretical models of disk instabilities produce eruptions that 
generally resemble FUor events, although comparisons with 
observations are still in their early stages (\cite{lin85}, \cite{cla89},
\cite{cla90}, \cite{kaw93}, \cite{bel94}, \cite{bel95}, \cite{cla96},
\cite{oku97}, \cite{tur97}).
The next few paragraphs summarize these results;
Herbig \& Petrov present a different interpretation
(\cite{her92}, \cite{pet98}; see also \cite{lar80}).

Figure 10 shows light curves for V1057 Cyg, one of the
best-studied FUors.  These light curves closely resemble
those of other accreting systems and provide strong support 
for an accretion model (\cite{ken88}, \cite{cla90}, \cite{ken91a}, 
\cite{bel95}, \cite{cla96}, \cite{lin96}, \cite{tur97}).  
The amplitude of the decline is largest in the UV and 
decreases monotonically from 0.3 $\mu$m to 5 $\mu$m.
This behavior is characteristic of accreting systems, which
evolve towards cooler temperatures as the luminosity declines
(\cite{ken84}, \cite{bel95}, \cite{lin96}, \cite{can98}).
The spectral type variation also agrees with disk model predictions.
If we assign stellar effective temperatures to the optical 
spectrum at maximum (A5 star) and at the current epoch (G5 star), 
the UBV decline indicates a source with a roughly constant radius.
This requirement is easy to achieve with an accretion disk
if the radius of the central star remains constant.  
Most stars cannot evolve at constant radius if their
effective temperatures change.

\epsfxsize=7.0in
\epsffile{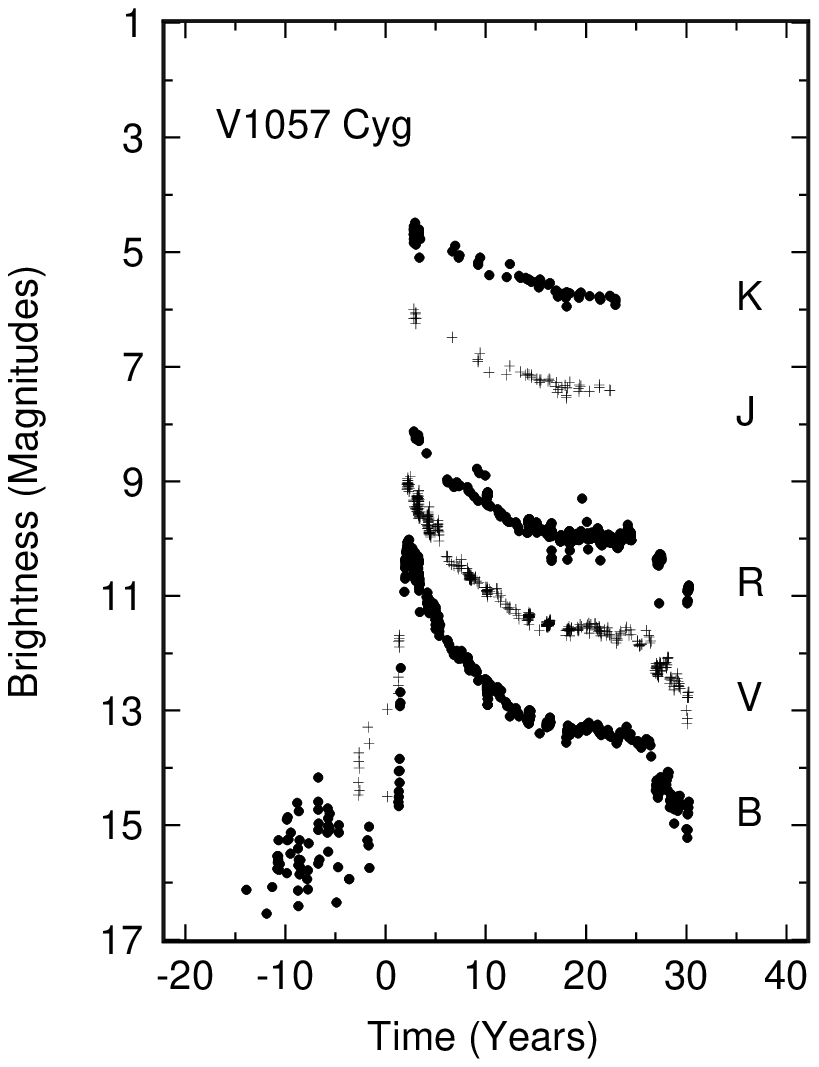}
\vskip -12ex

\footnotesize
\footspace

\noindent
Figure 10 - Light curves for V1057 Cyg.
The symbols alternate between filled circles (at B, R, and K)
and plus signs (V and J).
The system varied erratically at B = 14--17 prior to outburst
and then rose 5--6 mag in less than one year.
The visual observations indicate a similar rise time, although
few pre-outburst observations are available.
No pre-outburst near-IR data exist; K-band observations
began shortly after the outburst was detected.
The amplitude of the decline clearly depends on wavelength, 
with larger amplitudes at shorter wavelengths (see \cite{ken91a},
\cite{she95}, \cite{kol97}).

\vskip 3ex
\normalsize
\singlespace

The light curves of other FUors also generally agree with disk
model predictions.
The rise times of 200--400 days in FU Ori and V1057 Cyg are
comparable to the thermal time scale, as expected for an eruption
that begins at the inner edge of the disk (Table 3; \cite{bel95}).
The several decade rise in V1515 Cyg is too long for such an
``inside-out'' instability, but could be caused by an outburst
that begins in the outer part of the disk .  These ``outside-in''
eruptions can result from perturbations of the disk by a companion
star or planet, in addition to the disk instability mechanism
described above (\cite{cla90}, \cite{bon92}, \cite{bel95}, \cite{cla96}).  
In all systems, the decay times are comparable
to the viscous time scale (Table 3).

\epsfxsize=6.0in
\epsffile{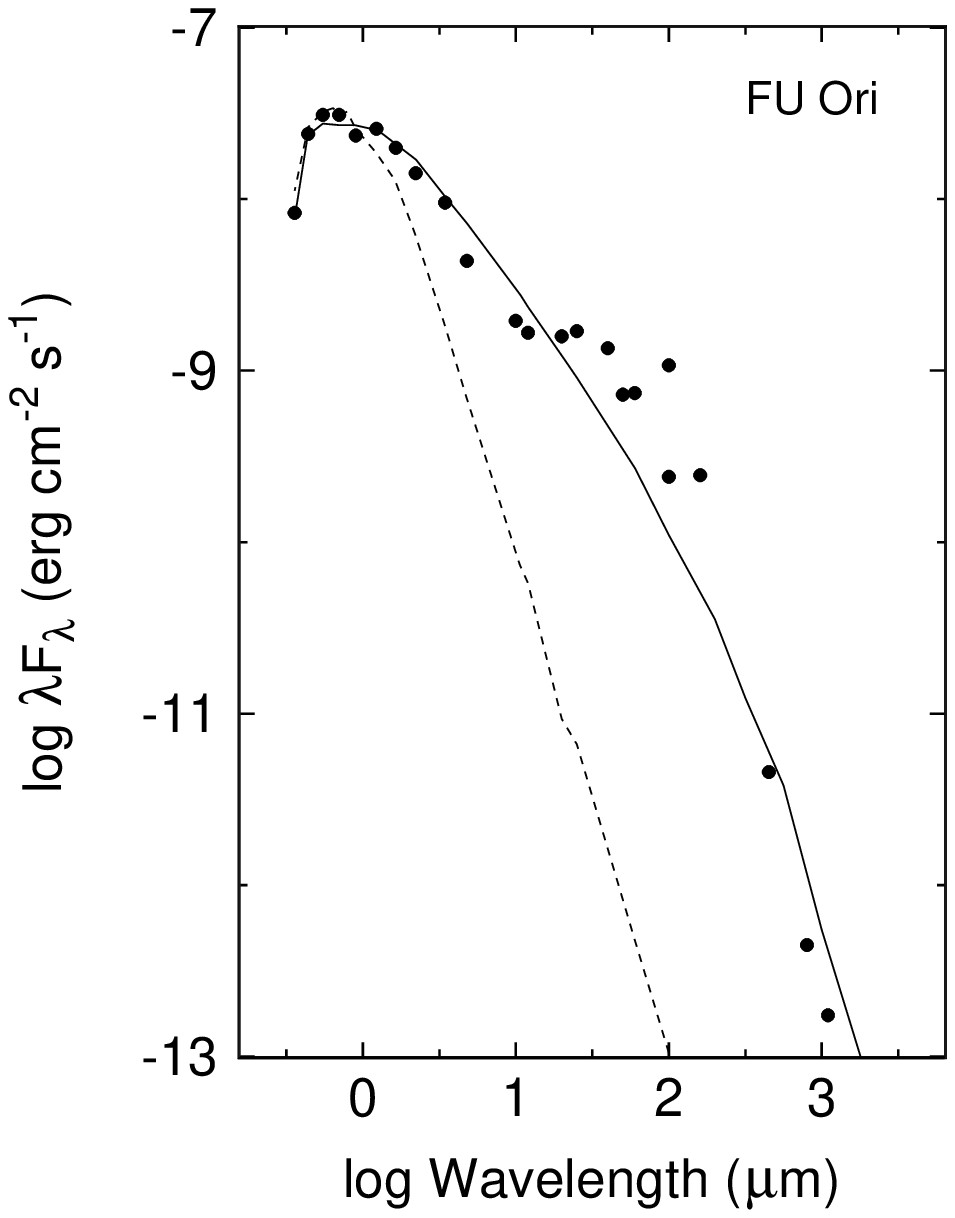}

\footnotesize
\footspace

\vskip -2ex
\noindent
Figure 11 -- Spectral energy distribution for FU Ori.  
The observations (filled circles) indicate a clear excess
over a G-type star (dashed line).  A disk model that produces
a G-type optical spectrum accounts for most of the data (solid line).
The modest far-IR excess over the disk spectrum at $\sim$ 100 \mum~is 
probably produced by the surrounding nebula responsible for the 
fan-shaped reflection nebula.
\vskip 3ex

\normalsize
\singlespace

Disk models also successfully explain FUor SEDs
(\cite{har85}, \cite{ken88}, \cite{ken91a}).
Figure 11 shows a dereddened SED for FU Ori.
The deep UV and near-IR absorption features indicate that
the UV and IR excesses over a G-type supergiant spectrum
are photospheric (\cite{cal91}).
If disks radiate like stars at the effective temperatures 
appropriate for the observed absorption features, the SEDs 
for FU Ori and other FUors require that the surface area 
of the emitting region increases with increasing wavelength.
In particular, the surface area of the near-IR source must be
10--20 times larger than the surface area of the optical source;
the 300 K material responsible for the 10 $\mu$m excess 
must have roughly 100 times the emitting area of the optical 
continuum region.
A disk in which the temperature decreases radially outward (\S2)
naturally explains this observation.
Disk models account for the 0.3--10 $\mu$m SEDs of V1057 Cyg,
V1515 Cyg, and FU Ori quite well (Figure 11; see also \cite{har85},
\cite{ada87}, \cite{ken88}, \cite{ken91a}, \cite{bel95}, \cite{tur97}).
Several other FUors 
with modest reddening have similar SEDs (\cite{ken91a}).

Recent interferometric observations have resolved FU Ori at
near-IR wavelengths. The angular size at 2.2 \mum~agrees with 
predictions of the accretion disk model (\cite{mal98}).

The disk model {\it fails} to explain 10--100 $\mu$m data
of many FUors, although the model fits both FU Ori and BBW 76
from 1--100 $\mu$m (\cite{ken91a}).
However, the decline of the 10--20 $\mu$m light of V1057 Cyg follows 
the optical light curve very closely (\cite{ken91a}).
This behavior suggests that the mid-IR radiation is optical light 
absorbed and reradiated by a surrounding envelope.
An infall rate of 1--5 $\times~10^{-6}$ \msunyr~produces 
an optically thick envelope that can reprocess optical light 
from the inner disk and account for the 10--100 $\mu$m SED
(\cite{ken91a}, \cite{bel95}).  
This rate is sufficient to replenish the disk in 1000 yr,
which allows recurrent FUor eruptions in this system.
The observational evidence for infall at comparable rates 
in two FUors, Z CMa (\cite{lil97}) and L1551 IRS5 (\cite{oha96},
\cite{mar97}), lends support to this interpretation.

Finally, an accretion disk also naturally explains the 
optical and IR line profiles observed in some FUors
(\cite{har87a}, \cite{har87b}, \cite{ken88}, \cite{har88},
\cite{wel91}, \cite{wel92}).  
The gradual decrease in the rotational velocity with increasing
wavelength occurs because the longer wavelength emission is 
produced in more slowly-orbiting material at larger disk radii
than the more rapidly moving inner disk material responsible
for the short wavelength emission.
At a given disk temperature, a larger fraction of 
the disk surface rotates at high line-of-sight velocities and 
the lines appear doubled (\cite{ken88}, \cite{cal93}, \cite{pop96}).

\subsection{The Importance of FU Ori Eruptions}

Despite their relative rarity, FUors are an important
part of pre-main sequence stellar evolution.  FUor
eruptions add a significant fraction of a stellar
mass to the central star.  FUor eruptions also eject
a large amount of mass into the surrounding cloud.
FUor eruptions may be frequent enough to provide 
nearly all of the mass of a typical low mass star.
If this hypothesis is correct, FUor eruptions can
power most molecular outflows and may be a dominant 
source of cloud heating in molecular cloud cores.  
This view is controversial, but can be tested with 
observations as outlined below.

Although the statistics are still crude, young stars
clearly accrete much of a stellar mass in FUor events.
A typical eruption adds $\sim$ 0.01 \msun~to the
central star; a rate of 10--20 eruptions per pre-main 
sequence star leads to a total accreted mass of $\sim$
0.1--0.2 \msun.  These eruptions eject considerable 
amounts of mass and momentum into the surrounding cloud.
For example, FU Ori itself has lost material with a 
momentum of $\sim$ 0.3 \msun~\kms~since its eruption 
began.  With 10--20 such eruptions, FU Ori could supply
a small fraction of the momentum of a typical 
molecular outflow, $(Mv)_o~\sim$ 1--100 \msun~\kms~(\cite{fuk89}).

Two conditions must be satisfied for a 
young star to accrete {\it all} of its mass in FUor events:
1. The disk must undergo regular thermal instabilities 
and cycle between the low and high states on a reasonable
time scale.  
2. FUors must be young, $\simless$ a few $\times~10^5$ yr, 
to allow the surrounding cloud to fuel the recurring instabilities.

As a group, FUors clearly are young objects.  FUors have 
many more properties in common with young class I protostars 
than older optically visible T Tauri stars.  FUors have 
distinctive reflection nebulae and large far-IR excesses; 
most are associated with jets, HH objects, and molecular outflows.
These properties are common in class I sources and rare in 
T Tauri stars (\cite{rei85}, \cite{rei90}, \cite{gom97}). 
The typical age of a FUor is therefore $\sim$ a 
few $\times~10^5$ yr (\cite{ken95}, \cite{wei91}, \cite{ken91a}).

The conditions needed for the disk to cycle between the
high and low states are probably satisfied in many pre-main
sequence stars.  For a typical infall rate of 1--10 $\times$
$10^{-6}~\msunyr$, theoretical models indicate that the disk 
spends most of its time in a low state where the mass accretion 
rate through the disk is much lower than the infall rate 
(\cite{bel94}, \cite{bel95}, \cite{lin96}).  The disk requires 
only $\sim$ $10^3$--$10^4$ yr to accumulate the $\sim$ 
0.01 \msun~needed to power a FUor eruption.  The disk will
continue to cycle between the low and high states as long as
the cloud supplies mass to the disk.  

If young stars accrete most of their mass in FUor events,
it follows that FUors can power molecular outflows. In most
outflow models, the mass ejected by the central star is 
$\sim$ 10\%--30\% of the accreted mass. For a wind
velocity of $\sim$ 200~\kms, the momentum in the wind is

\begin{equation}
(Mv)_w ~ \sim ~ 20 ~ (\mstar/\msun) ~ \msun~\kms
\end{equation}

\noindent
for a typical wind velocity of 200 \kms.  The observed range of
outflow momenta -- $(Mv)_o$ = 1--100 \msun~\kms~(\cite{fuk89}) --
requires stellar masses of $\mstar ~ \sim$ 0.1--5 \msun~if
the flows conserve momentum.  The winds of FUors can power this
outflow {\bf if} a young star accretes most of its mass in FUor events.

These simple estimates establish that FUors are important events
if they recur on time scales of $\sim 10^3$ yr during the protostellar
phase of pre-main sequence stellar evolution.  In this picture,
the fraction of time in the FUor state is simply the 
ratio of the infall rate to the FUor accretion rate: 
$f \sim \mdot_i/\mdot_{FU}$.  FUor disk models require
$\mdot_{FU} \sim 10^{-4}$ \msunyr, so $f \sim 0.05$ for a 
typical infall rate of $\mdot_i \sim 5 \times 10^{-6}$ \msunyr.
This prediction currently agrees with the statistics of FUors
among protostars: 1 out of $\sim$ 20 protostars in
the Taurus dark cloud contains a FUor (L1551 IRS5; \cite{ken95})
and 5 out of 17 protostars with HH objects contains a
central star that resembles known FUors spectroscopically
(\cite{re97b}; see also \cite{han95}).  Sensitive photometric and 
spectroscopic surveys 
of nearby molecular clouds can improve these statistics considerably.
If the FUor frequency among protostars turns out to be more 
than a few per cent, then FUors may represent the main accretion 
phase of early stellar evolution.

I thank Janet Mattei for the SS Cyg data and John Cannizzo for
the disk instability models.


\begin{thebibliography}{200}

\bibitem[Adams \etal 1987]{ada87} Adams, F. C., Lada, C. J., \& Shu, F. H. 
1987, ApJ, 308, 788

\bibitem[Alves \etal 1997]{alv97} Alves, J., Hartmann, L., Brice\~no, C.,
\& Lada, C. J. 1997, AJ, 113, 1395

\bibitem[Ambartsumian 1954]{amb54} Ambartsumian, V. A. 1954, 
Comm. Byurakan Obs, No 13

\bibitem[Armitage 1995]{arm95} Armitage, P. J. 1995, MNRAS, 274, 1242

\bibitem[Armitage 1998]{arm98} Armitage, P. J. 1998, ApJL, 501, L189.

\bibitem[Bachiller \etal 1994]{bac94} Bachiller, R., Tafalla, M., \&
Cernicharo, J. 1994, ApJo, 425, L93

\bibitem[Balbus \& Hawley 1991]{bal91} Balbus, S. A., \& Hawley, J. F.
1991, ApJ, 376, 214

\bibitem[Balbus \& Hawley 1998]{bal98} Balbus, S. A., \& Hawley, J. F.
1998, Rev Mod Phys, 70, 1

\bibitem[Balbus \etal 1996]{bal96} Balbus, S. A., Hawley, J. F., \&
Stone, J. M.  1996, ApJ, 467, 76

\bibitem[Baptista \etal 1998]{bap98} Baptista, R., Horne, K., Wade, R. A.,
Hubeny, I., Long, K. S., \& Rutten, R. G. M. 1998, MNRAS, 298, 1079

\bibitem[Barth \etal 1994]{bar94} Barth, W., Weigelt, H., \&
Zinnecker, H. 1994, A\&A, 291, 500


\bibitem[Bastian \& Mundt 1985]{bas85} Bastian, U., \& Mundt, R. 1985, 
A\&A, 144, 57

\bibitem[Bath \etal 1974]{bat74} Bath, G.T., Evans, W.D., 
Papaloizou, J., \& Pringle, J.E. 1974, MNRAS, 169, 447


\bibitem[Bell \& Lin 1994]{bel94} Bell, K. R., \& Lin, D. N. C. 1994,
ApJ, 427, 987

\bibitem[Bell \etal 1995]{bel95} Bell, K. R., Lin, D. N. C., Hartmann, L., \& 
Kenyon, S. J. 1995, ApJ, 444, 376

\bibitem[Belyakina 1992]{bel92} Belyakina, T. S. 1992, Izv. Krym. Ap. Obs.,
84, 45



\bibitem[Bertout \etal 1993]{ber93} Bertout, C., Bouvier, J., Duschl, W.J.,
\& Tscharnuter, W.M. 1993, A\&A, 275, 236

\bibitem[Bertout \& Regev 1992]{ber92} Bertout, C., \& Regev, O. 
1992, ApJ, 399, L163

\bibitem[Biscaya \etal 1997]{bis97} Biscaya, A. M., Rieke, G. H., 
Narayanan, G., Luhman, K. L., \& Young, E. T. 1997, AJ, 491, 359

\bibitem[Bonnell \& Bastian 1992]{bon92} Bonnell, I., \& Bastian, P. 1992,
ApJ, 401, L31

\bibitem[Bouvier \& Bertout 1989]{bou89} Bouvier, J., \& Bertout, 1989, A\&A, 211, 99

\bibitem[Bouvier \etal 1993]{bou93} Bouvier, J., Cabrit, S., 
Fernandez, M., Martin, E. L., \& Matthews, J. M. 1993, A\&A, 272, 176

\bibitem[Bruch 1994]{bru94} Bruch, A. 1994, in
{\it Flares and Flashes}, IAU Colloquium No. 151, 
edited by J. Greiner, H. W. Duerbeck, \& R. E. Gershberg,
Berlin, Springer, p. 288

\bibitem[Cabot 1996]{cab96} Cabot, W. 1996, ApJ, 465, 874

\bibitem[Cabot \etal 1987]{cab87} Cabot, W., Canuto, V. M., 
Hubickyj, O., \& Pollack, J. B. 1987, Icarus, 69, 387

\bibitem[Calvet \& Gullbring 1999]{cal99} Calvet, N. \& Gullbring, E. 1999,
ApJ, in press

\bibitem[Calvet \etal 1991]{cal91} Calvet, N., Hartmann, L., \& Kenyon, S. J. 
1991, ApJ, 383, 752

\bibitem[Calvet \etal 1993]{cal93} Calvet, N., Hartmann, L., \& Kenyon, S. J. 
1993, ApJ, 402, 623

\bibitem[Campbell 1987]{cam87} Campbell, C. G. 1987, MNRAS, 229, 405

\bibitem[Cannizzo \& Mattei 1998]{can98} Cannizzo, J. K., \& Mattei, J. A.
1998, ApJ, 505, 344

\bibitem[Carr \etal 1987]{car87} Carr, J. S., Harvey, P. M., \& Lester, D. F. 
1987, ApJL, 321, L71


\bibitem[Clarke \etal 1989]{cla89} Clarke, C. J., Lin, D. N. C., \& 
Papaloizou, J. C. B.  1989, MNRAS, 236, 495

\bibitem[Clarke \etal 1990]{cla90} Clarke, C. J., Lin, D. N. C., \& 
Pringle, J. E. 1990, MNRAS, 242, 439

\bibitem[Clarke \& Syer 1996]{cla96} Clarke, C. J., \& Syer, D. 1996,
MNRAS, 278, L23

\bibitem[Cohen \etal 1982]{coh82} Cohen, M., Beiging, J. H., \& 
Schwartz, P. R. 1982, ApJL, 289, L5

\bibitem[Cropper 1990]{cro90} Cropper, M. 1990, Sp. Sci. Rev., 54, 195

\bibitem[Croswell \etal 1987]{cro87} Croswell, K., 
Hartmann, L., \& Avrett, E. 1987, ApJ, 312, 227

\bibitem[Davis \etal 1994]{dav94} Davis, C. J., Mundt, R.,
Eisl\"offel, J., \& Ray, T. P. 1994, AJ, 110, 766


\bibitem[Eisl\"offel \etal 1990]{eis90} Eisl\"offel, J., Hessman, F. V.,
\& Mundt, R. 1990, A\&A, 232, 70
\bibitem[Elias 1978]{eli78} Elias, J. H. 1978, ApJ, 223, 859

\bibitem[Eisl\"offel \etal 1991]{eis91} Eisl\"offel, J., G\"unther, E.,
Hessman, F. V., Mundt, R., Carr, J. S., Beckwith, S., Ray, T. P. 1991,
ApJ, 383, L19

\bibitem[Evans \etal 1994]{eva94}
Evans, II, N. J., Balkum, S., Levreault, R. M., Hartmann, L., 
\& Kenyon, S. J. 1994, ApJ, 424, 793

\bibitem[Fridlund \& Knee 1993]{fri93} Fridlund, C. V. M., Knee, L. B. G. 1993,
A\&A, 268, 245

\bibitem[Fridlund \& Liseau 1998]{fri98} Fridlund, C. V. M., \& Liseau 1998,
ApJ, 499, L75

\bibitem[Fukui 1989]{fuk89} Fukui, Y. 1989, in 
{\it ESO Workshop on Low Mass Star Formation and 
Pre-Main Sequence Objects,} edited by B. Reipurth, Garching, ESO, p. 95

\bibitem[Ghosh \& Lamb 1979]{gho79} Ghosh, P., \& Lamb, F. K. 1979, ApJ, 232, 259

\bibitem[Godon \etal 1995]{god95} Godon, P., Regev, O., \& Shaviv, G. 1995,
MNRAS, 275, 1093

\bibitem[Godon 1996]{god96} Godon, P. 1996, MNRAS, 279, 1071

\bibitem[G\'omez \etal 1997]{gom97} G\'omez, M., Whitney, B. A., \&
Kenyon, S. J. 1997, AJ, 114, 1138

\bibitem[Goodrich 1987]{goo87} Goodrich, R. 1987, PASP, 99, 116

\bibitem[Greene \& Lada 1997]{gre97} Greene, T. P., \& Lada, C. J. 
1997, AJ, 114, 2157

\bibitem[Hanson \& Conti 1995]{han95} Hanson, M. M., \& Conti, P. S.
1995, ApJ, 448, L45

\bibitem[Hartigan \etal 1990]{hrt90} Hartigan, P., Raymond, J., \& Meaburn, J.
1990, ApJ, 362, 624

\bibitem[Hartmann \& Calvet 1995]{har95} Hartmann, L., \& Calvet, N. 1995,
AJ, 109, 1846

\bibitem[Hartmann \& Kenyon 1985]{har85} Hartmann, L., \& Kenyon, S.J. 
1985, ApJ, 299, 462

\bibitem[Hartmann \& Kenyon 1987a]{har87a} Hartmann, L., \& Kenyon, S.J. 
1987a, ApJ, 312, 243.

\bibitem[Hartmann \& Kenyon 1987b]{har87b} Hartmann, L., \& Kenyon, S.J. 
1987b, ApJ, 322, 393

\bibitem[Hartmann \etal 1988]{har88} Hartmann, L., Kenyon, S. J., Hewett, R., 
Edwards, S., Strom, K. M., Strom, S. E., \& Stauffer, J. R. 1988, ApJ, 338, 1001

\bibitem[Hartmann \& Kenyon 1996]{har96} Hartmann, L., \& Kenyon, S. J. 
1996, ARA\&A,

\bibitem[Hawley \& Balbus 1991]{haw91} Hawley, J. F., \& Balbus, S. A. 
1991, ApJ, 376, 223


\bibitem[Herbig 1960]{her60} Herbig, G. H. 1960, ApJS, 4, 33

\bibitem[Herbig 1966]{her66} Herbig, G. H. 1966, Vistas in Astr, 8, 109

\bibitem[Herbig 1977]{her77} Herbig, G. H. 1977, ApJ, 

\bibitem[Herbig 1989]{her89} Herbig, G. H. 1989, in
\it ESO Workshop on Low-Mass Star Formation and Pre-Main Sequence Objects \rm
ed. B. Reipurth, Garching, ESO, p. 233

\bibitem[Herbig \& Petrov 1992]{her92} Herbig, G. H., \& Petrov, P. P. 
1992, ApJ, 392, 209

\bibitem[Herbst \etal 1987]{hrb87} Herbst, W., \etal 1987, AJ, 94, 137

\bibitem[Herbst \etal 1994]{hrb94} Herbst, W., Herbst, D. K., 
\& Grossman, E. J. 1994, AJ, 108, 1906

\bibitem[Herrnstein \etal 1998]{her98} Herrnstein, J. R., Greenhill, L. J.,
Moran, J. M., Diamond, P. J., Inque, M., Nakai, N., Miyoshi, M. 1998,
ApJ, 497, L69

\bibitem[Hodapp \etal 1996]{hod96} Hodapp, K.-W., Hora, J. L.,
Rayner, J. T., Pickles, A. J., \& Ladd, E. F. 1996, ApJ, 468, 861

\bibitem[Horne \& Cook 1985]{ho85a} Horne, K., \& Cook, M. C.
1985, MNRAS, 214, 307

\bibitem[Horne \& Steining 1985]{ho85b} Horne, K., \& Steining, R. F. 
1985, MNRAS, 216, 933

\bibitem[Hujeirat 1995a]{hu95a} Hujeirat, A. 1995a, A\&A, 295, 249

\bibitem[Hujeirat 1995b]{hu95b} Hujeirat, A. 1995b, A\&A, 295, 268

\bibitem[Ibragimov 1993]{ibr93} Ibragimov, M. A. 1993, Astr. Zh., 70, 339

\bibitem[Ibragimov 1997]{ibr97} Ibragimov, M. A. 1997, Pis'ma Astr. Zh., 
1, 125

\bibitem[Idan \& Shaviv 1996]{ida96} Idan, I. \& Shaviv, G. 
1996, MNRAS, 281, 604 

\bibitem[Jayawardhana \etal 1998]{jay98} Jayawardhana, R., Fisher, S.,
Hartmann, L., Telesco, C., Pina, R., \& Fazio, G. 1998, ApJ:, 503, L79

\bibitem[Joy 1945]{joy45} Joy, A. H. 1945, ApJ, 102, 168

\bibitem[Kant 1755]{kan55} Kant, I. 1755,
{\it Universal Natural History and Theories of the Heavens}

\bibitem[Kawazoe \& Mineshige 1993]{kaw93} Kawazoe, E., \& Mineshige, S.
1993, PASJ, 45, 715

\bibitem[Kenyon 1986]{ken86}
Kenyon, S. J. 1986, {\it The Symbiotic Stars,} Cambridge University Press

\bibitem[Kenyon \etal 1993a]{ken93a} Kenyon, S. J., Calvet, N., \& Hartmann, L. 1993a, ApJ, 414, 676

\bibitem[Kenyon \& Hartmann 1987]{ken87} Kenyon, S. J., \& Hartmann, L. 
1987, ApJ, 323, 714


\bibitem[Kenyon \& Hartmann 1991]{ken91a} Kenyon, S. J., \& Hartmann, L. 
1991, ApJ, 383, 664

\bibitem[Kenyon \& Hartmann 1995]{ken95} Kenyon, S. J., \& Hartmann, L. 
1995, ApJS, 101, 117

\bibitem[Kenyon \etal 1993b]{ken93b} Kenyon, S. J., Hartmann, L., G\'omez, M., 
Carr, J., \& Tokunaga, A.  1993b, AJ, 105, 1505

\bibitem[Kenyon \etal 1988]{ken88} Kenyon, S. J., Hartmann, L., \& 
Hewett, R. 1988, ApJ, 325, 231

\bibitem[Kenyon \etal 1994]{ken94} Kenyon, S. J., \etal 1994, AJ, 107, 2153

\bibitem[Kenyon \etal 1989]{ke89b} Kenyon, S.J., Hartmann, L.W., 
Imhoff, C.L., \& Cassatella, A. 1989, ApJ, 344, 925.

\bibitem[Kenyon \etal 1991]{ken91b} Kenyon, S. J., Hartmann, L., \& 
Kolotilov, E. A. 1991, PASP, 103, 1069

\bibitem[Kenyon \& Webbink 1984]{ken84} 
Kenyon, S. J., \& Webbink, R. F. 1984, ApJ, 279, 252

\bibitem[Kenyon \etal 1993c]{ken93c} Kenyon, S. J., Whitney, B., 
G\'omez, M., \& Hartmann, L.  1993c, ApJ, 414, 773

\bibitem[Kenyon \etal 1996]{ken96} Kenyon, S. J., Yi, I.  \& Hartmann, L.
1996, ApJ, 462, 439

\bibitem[Kholopov 1959]{kho59} Kholopov, P. N. 1959, Sov. AJ, 3, 291

\bibitem[Kley 1991]{kle91} Kley, W. 1991, A\&A, 247, 95

\bibitem[Kley \& Hensler 1987]{kle87} Kley, W., \& Hensler, G. 1987, 
A\&A, 172, 124

\bibitem[Kley \& Lin 1996]{kle96} Kley, W., \& Lin, D. N. C. 1996, 
ApJ, 461, 933

\bibitem[Kley \& Papaloizou 1997]{kle97} Kley, W., \& Papaloizou, J. C. B.
1997, MNRAS, 285, 239

\bibitem[Kolotilov \& Kenyon 1997]{kol97} Kolotilov, E. A., 
\& Kenyon, S. J. 1998, IBVS, No. 4494

\bibitem[Kolotilov \& Petrov 1983]{kol83} Kolotilov, E. A., 
\& Petrov, P. P. 1983, Pis'ma Astr. Zh., 9, 171

\bibitem[K\"{o}nigl 1991]{kon91} K\"{o}nigl, A. 1991, ApJL, 370, L39

\bibitem[Koresko \etal 1991]{kor91} Koresko, C. D., Beckwith, S. V. W., 
Ghez, A. M., Matthews, K., Neugebauer, G. 1991, AJ, 102, 2073

\bibitem[Kuiper 1941]{kui41} Kuiper, G. P. 1941, ApJ, 93, 133

\bibitem[Lamzin 1998]{lam98} Lamzin, S. A. 1998, Astr. Rept., in press

\bibitem[Laplace 1796]{lap96} Laplace, P. S. 1796, {\it M\'ecanique C\'eleste}

\bibitem[Larson 1980]{lar80} Larson, R. B. 1980, MNRAS, 190, 321

\bibitem[Li 1996]{li96} Li, J. 1996, ApJ, 456, 696

\bibitem[Liljestr\"om \& Olofsson 1997]{lil97} Liljestr\"om, T., \& 
Olofsson, G. 1997, ApJ, 478, 381

\bibitem[Lin \& Papaloizou 1985]{lin85}
Lin, D. N. C., \& Papaloizou, J. 1985, in
{\it Protostars and Planets II,}
ed. D. C. Black and M. S. Matthews,
Tucson, University of Arizona Press, p. 981

\bibitem[Lin \& Papaloizou 1995]{lin95}  Lin, D. N. C., \& Papaloizou, J. C. B.,
1996, ARA\&A, 33, 505

\bibitem[Lin \& Papaloizou 1996]{lin96} Lin, D. N. C., \& Papaloizou, J. C. B.,
1996, ARA\&A, 34, 703

\bibitem[Lin \& Pringle 1987]{lin87} Lin, D. N. C., \& Pringle, J. E.,
1987, MNRAS, 225, 607

\bibitem[Lipunov 1978]{lip78} Lipunov, V. M. 1978, Astron. Zh., 55, 1233

\bibitem[Lipunov 1980]{lip80} Lipunov, V. M. 1980, Astron. Zh., 57, 1253

\bibitem[Livio 1997]{liv97} Livio, M. 1997, in ASP Conf. Ser 121,
{\it Accretion Phenomena and Related Outflows,} edited by 
D. T. Wickramasinghe, G. V. Bicknell, \& L. Ferrario,
San Francisco, ASP, p. 845

\bibitem[Looney \etal 1997]{loo97} Looney, L. W., Mundy, L. G., \& 
Welch, W. J. 1997, ApJ, 484, L157

\bibitem[L\'opez \etal 1998]{lop98} L\'opez, R., \etal 1998, AJ, 116, 845

\bibitem[Lucas \& Roche 1996]{luc96} Lucas, P. W., \& Roche, P. F. 1996,
MNRAS, 280, 1219

\bibitem[L\"ust 1952]{lus52} L\"ust, R. 1952, Zs. f. Nat., 7a, 87

\bibitem[Lynden-Bell \& Pringle 1974]{lyn74} Lynden-Bell, D., \& 
Pringle, J. E. 1974, MNRAS, 168, 603

\bibitem[McMuldroch \etal 1995]{mcm95} McMuldroch, S., Blake, G. A., \&
Sargent, A. I. 1995, AJ, 110, 354

\bibitem[Mahdavi \& Kenyon 1998]{mah98} Mahdavi, A., \& Kenyon, S. J. 
1998, ApJ, 497, 342

\bibitem[Malbet \etal 1998]{mal98} Malbet, F., \etal 1998, ApJL, 507, 149

\bibitem[Mardones \etal 1997]{mar97} Mardones, D., Myers, P. C., 
Tafalla, M., Wilner, D. J., Bachiller, R., \& Garay, G. 1997, ApJ, 489, 719

\bibitem[Marsh \& Horne 1988]{mar88} Marsh, T. R., \& Horne, K. 1988,
MNRAS, 235, 269

\bibitem[Mould \etal 1978]{mou78} Mould, J. R., Hall, D. N. B., 
Ridgway, S. T., Hintzen, P., \& Aaronson, M. 1978, ApJL, 222, L123

\bibitem[Nakajima \& Golimowski 1995]{nak95} Nakajima, T., \& Golimowski, D. A.
1995, AJ, 109, 1181

\bibitem[Ohashi \etal 1996]{oha96} Ohashi, N., Hayashi, M., Ho, P.T.P.,
Momose, M., \& Hirano, N. 1996, ApJ, 466, 957

\bibitem[Okuda \etal 1997]{oku97} Okuda, T., Fujita, M., \&
Sakashita, S. 1997, PASJ, 49, 679

\bibitem[Oppenheimer \etal 19]{opp98} Oppenheimer, B. D., Kenyon, S. J.,
\& Mattei, J. A. 1998, AJ, 115, 1175

\bibitem[Osaki 1974]{osa74} Osaki, Y. 1974, PASJ, 26, 429

\bibitem[Papaloizou \& Stanley 1986]{pap86} Papaloizou, J. C. B., \& 
Stanley, G. Q. G. 1986, MNRAS, 220, 593

\bibitem[Petrov \etal 1998]{pet98} Petrov, P., Duemmler, R., 
Ilyin, I., Tuominen, I. 1998, A\&A, 331, L53

\bibitem[Popham \& Narayan 1991a]{pop91a} Popham, R., \&  Narayan, R. 
1991, ApJ, 370, 604

\bibitem[Popham \& Narayan 1991b]{pop91b} Popham, R., \&  Narayan, R. 
1991, ApJ, 394, 255

\bibitem[Popham \etal 1993]{pop93} Popham, R., Narayan, R., 
Hartmann, L., \& Kenyon, S. J.  1993, ApJL, 415, L127

\bibitem[Popham \etal 1996]{pop96} Popham, R., Narayan, R., 
Kenyon, S. J., \& Hartmann, L. 1996, ApJ, 473, 422

\bibitem[Pringle 1977]{pri77} Pringle, J. E. 1977, MNRAS, 178, 95

\bibitem[Pringle \etal 1973]{pri73} Pringle, J. E., Rees, M. J.,
\& Pacholczyk, A. G. 1973, A\&A, 29, 179

\bibitem[Pringle \& Savonije 1979]{pri79} Pringle, J. E., \& Savonije, G. J. 
1979, MNRAS, 187, 777

\bibitem[Regev 1983]{reg83} Regev, O. 1983, A\&A, 126, 146.

\bibitem[Regev \& Bertout 1995]{reg95} Regev, O., \& Bertout, C.  
1995, MNRAS, 272, 71

\bibitem[Regev \& Hougerat 1988]{reg88} Regev, O., \& Hougerat, A. A. 
1988, MNRAS, 232, 81

\bibitem[Reg\"os 1997]{reg97} Reg\"os, E. 1997, MNRAS, 286, 104

\bibitem[Reipurth 1985]{rei85} Reipurth, B. 1985, A\&A, 143, 435

\bibitem[Reipurth 1989]{rei89} Reipurth, B. 1989, Nature, 340, 42

\bibitem[Reipurth 1990]{rei90} Reipurth, B. 1990, in
{\it Flare Stars in Star Clusters, Associations, and the Solar Vicinity,}
IAU Symposium No. 137, Dordrecht, Kluwer, p. 229

\bibitem[Reipurth 1991]{rei91} Reipurth, B. 1991, in
{\it Physics of Star Formation and Early Stellar Evolution},
NATO Adv. Study Inst., edited by C. J. Lada and N. D. Kylafis, p. 497

\bibitem[Reipurth 1997]{re97a} Reipurth, B. 1997, in
{\it Low Mass Star Formation -- from Infall to Outflow,} 
Poster Proceedings of IAU Symposium No. 182 on Herbig-Haro Objects 
and the Birth of Low Mass Stars, edited by F. Malbet \& A. Castets, p. 309

\bibitem[Reipurth \& Aspin 1997]{re97b} Reipurth, B., \& Aspin, C. 
1997, AJ, 114, 2700

\bibitem[Reipurth \etal 1997]{re97c} Reipurth, B., Bally, J., \& Devine, D. 
1997, AJ, 114, 2708

\bibitem[Reipurth \& Heathcote 1992]{rei92} Reipurth, B., \& Heathcote, S.
1992, A\&A, 257, 693

\bibitem[Reipurth \etal 1997]{re97d} Reipurth, B., Olberg, M., Gredel, R., 
\& Booth, R. S. 1997, A\&A, 327, 1164

\bibitem[Rodriguez \etal 1990]{rod90} Rodriguez, L., Hartmann, L. W., 
\& Chavira, E. 1990, PASP, 102, 1413

\bibitem[Rodriguez \& Hartmann 1992]{rod92} Rodriguez, L., \& Hartmann, L. 
1992, Rev. Mex. A\&A, 24, 135

\bibitem[Rodriguez \etal 1998]{rod98} Rodriguez, L., D'Alessio, P., 
Wilner, D.J., Ho, P.T.P., Torrelles, J.M., Curiel, S., G\'omez, Y.,
Lizano, S., Pedlar, A., Canto, J., \& Raga, A.C. 1998, Nat, 395, 355

\bibitem[Rozyczka \etal 1996]{roz96} Rozyczka, M., Bodenheimer, P., \&
Lin, D. N. C. 1996, ApJ, 459, 371

\bibitem[Rucinski 1985]{ruc85} Rucinski, S. M. 1985, AJ, 90, 2321

\bibitem[Sandell \& Aspin 1998]{san98} Sandell, G., \& aspin, C.
1998, A\&A, 333, 1016

\bibitem[Sato \etal 1992]{sat92} Sato, S., Okita, K., Yayamshita, T.,
Mizutani, K., Shiba, H., Kobayashi, Y., \& Takami, H. 1992, ApJ, 398, 273

\bibitem[Shakura \& Sunyaev 1973]{sha73} Shakura, N. I., 
\& Sunyaev, R. A. 1973, A\&A, 24, 337

\bibitem[Shevchenko 1995]{she95} Shevchenko, V. S. 1995, rotor project

\bibitem[Shevchenko \etal 1997]{she97} Shevchenko, V. S., 
Ezhkova, O., Tjin A Djie, H. R. E., van den Anckner, M. E., 
Blondel, P. F. C., \&  de Winter, D. 1997, A\&AS, 124, 33

\bibitem[Smith \& Terrile 1984]{smi84} Smith, B. A., \& Terrile, R. J. 
1984, Science, 226, 1421

\bibitem[Staude \& Neckel 1991]{sta91} Staude, H. J., \& Neckel, Th. 
1991, A\&A, 244, L13

\bibitem[Staude \& Neckel 1992]{sta92} Staude, H. J., \& Neckel, Th. 
1992, ApJ, 400, 556

\bibitem[Stocke \etal 1988]{sto88} Stocke, J. T., Hartigan, P. M., 
Strom, S. E., Strom, K. M., Anderson, E. R., Hartmann, L. W., \& 
Kenyon, S. J. 1988, ApJS, 68, 229

\bibitem[Stone \& Balbus 1996]{sto96} Stone, J. M. \& Balbus, S. A. 
1996, ApJ, 464, 364

\bibitem[Strom \& Strom 1993]{str93}Strom, K. M., \& Strom, S. E. 
1993, ApJL, 421, L63

\bibitem[Teodorani \etal 1997]{teo97} Teodorani, M., Errico, L., 
Vittone, A. A., Giovanelli, F., \& Rossi, C. 1997, A\&AS, 126, 91

\bibitem[Thiebaut \etal 1995]{thi95} Thiebaut, E., Bouvier, J., Blazit, A.,
Bonneau, D., Foy, F.-C. \& Foy, R. 1995, A\&A, 303, 795

\bibitem[Tout \& Pringle 1992]{tou92} Tout, C. A., \& Pringle, J. E.
1992, MNRAS, 259, 604

\bibitem[Turner \etal 1997]{tur97} Turner, N. J. J., Bodenheimer, P.,
\& Bell, K. R. 1997, ApJ, 480, 754

\bibitem[Tylenda 1981]{tyl81} Tylenda, R. 1981, Acta Astr., 31, 267

\bibitem[Ulrich \etal 1997]{ulr97} Ulrich, M.-H., Maraschi, L., \& 
Urry, C. M. 1997, ARA\&A, 35, 445

\bibitem[Vishniac \& Diamond 1989]{vis89} Vishniac, E. T., \& 
Diamond, P. 1989, ApJ, 347, 435

\bibitem[Vishniac \& Zhang 1996]{vis96} Vishniac, E. T., \& 
Zhang, C. 1996, ApJ, 461, 307

\bibitem[von Weisz\"acker 1943]{von43} von Weisz\"acker, C. F. 1943,
Zs. f. Ap., 22, 319

\bibitem[von Weisz\"acker 1948]{von48} von Weisz\"acker, C. F. 1948,
Zs. f. Nat., 3a, 524

\bibitem[Wachman 1939]{wac39} Wachman, A. A. 1939, Beob. Zirk., 21, 60

\bibitem[Wachman 1954]{wac54} Wachman, A. A. 1954, Zs. f. Ap., 35, 74

\bibitem[Wang 1997]{wan97} Wang, Y. M. 1997, ApJL, 475, 135

\bibitem[Weintraub \etal 1989]{wei89} Weintraub, D. A., Sandell, G.,
\& Duncan, W. D. 1989, ApJ, 340, 69

\bibitem[Weintraub \etal 1991]{wei91} Weintraub, D. A., Sandell, G.,
\& Duncan, W. D. 1991, ApJ, 382, 270

\bibitem[Welin 1971]{wel71} Welin, G. 1971, A\&A, 12, 312

\bibitem[Welty \etal 1991]{wel91} Welty, A. D., Strom, S. E., Strom, K. M., 
Hartmann, L., Kenyon, S. J., Grasdalen, G. L., \& Stauffer, J. R. 1991, 
ApJ, 349, 328

\bibitem[Welty \etal 1992]{wel92} Welty, A. D., Strom, S. E., Edwards, S., 
Kenyon, S. J., \& Hartmann, L. W. 1992, ApJ, 397, 260

\bibitem[Whitney \etal 1993]{whi93} Whitney, B. A., Clayton, G. C., 
Schulte-Ladbeck, R. E., Calvet, N., Hartmann, L. \& Kenyon, S. J. 1993,
ApJ, 417, 687

\bibitem[Yamashita \& Tamura 1992]{yam92} Yamashita, T., \& Tamura, M. 1992,
ApJL, 387, L93

\bibitem[Yang \etal 1995]{yan95} Yang, J., Ohashi, N., \& Fukui, Y. 1995,
ApJ, 455, 175

\bibitem[Yi 1994]{yi94} Yi, I. 1994, ApJ, 428, 760

\bibitem[Yun \etal 1997]{yun97} Yun, J. L., Moreira, M. C., Alves, J. F.,
\& Storm, J. 1997, A\&A, 320, 167

\end{thebibliography}
\end{document}